\documentclass[11pt,a4paper]{article}
\usepackage{amsmath,amssymb,amsfonts,a4wide,graphicx,bm,times}
\usepackage{mcite,textcomp}
\usepackage[mathscr]{euscript}

\makeatletter \let\old@startsection=\@startsection
\renewcommand{\@startsection}[6]
{\old@startsection{#1}{#2}{#3}{#4}{#5}{#6\mathversion{bold}}}
\makeatother

\makeatletter

\@addtoreset{equation}{section}
\makeatother
\let\refOld\ref
\renewcommand{\ref}[1]{(\refOld{#1})}

\newcommand{\superp}[2]{\genfrac{}{}{0pt}{}{#1}{#2}}

 \def\d{\delta}

 \def\Im{{\rm Im ~}}
 \def\p{\partial}
 
 \def\a{\alpha}
 \def\b{\beta}
 
 \def\d{\delta}
 \def\e{\epsilon}
 \def\ve{\varepsilon}
 \def\th{\theta}

 \def\l{\lambda}

 \def\x{\xi}

 \def\s{\sigma}
 
 \def\th{\theta}

 \def\G{\Gamma}
 \def\D{\Delta}

 \def\L{\Lambda}
 \def\O{\Omega}
 \def\o{\omega }
 
\def\hf{\dfrac{1}{2}}

\def\lbrackdbl{[\![}
\def\rbrackdbl{]\!]}
\def\implies{\quad\Rightarrow\quad}

\def\vphi{\varphi}

\def\CF{\mathcal{F}}

\def\CZ{{\mathcal{Z}}}

\def\pert{^{(\text{pert})}}

\def\brho{\bar\rho}

\def\bC{\bar C}

\def\CS{\mathcal{S}}

\def\tl{\tilde{\l}}

\def\SGC{\CS_\text{GC}}

\def\ZGC{\CZ_\text{GC}}
\def\ZC{\CZ_\text{C}}
\def\FGC{\CF_\text{GC}}

\def\tl{\tilde{\l}}

\def\tt{\tilde{t}}
\def\Nc{N_c}
\def\NB{N_B}
\def\NBb{\bar{N}_B}
\def\bu{\bar{u}}
\def\bv{\bar{v}}
\def\bx{\bar{x}}
\def\tx{\tilde{x}}
\def\bk{\bar{k}}
\def\SJ{\mathscr{J}}

\begin{document}
\begin{titlepage}
\renewcommand{\thefootnote}{\fnsymbol{footnote}}
\vspace*{-2cm}
\begin{flushright}
APCTP Pre2014-010\\
\end{flushright}

\vspace*{1cm}
    \begin{Large}
       \begin{center}
         {\huge Spherical Hecke algebra in the Nekrasov-Shatashvili limit}
       \end{center}
    \end{Large}
\vspace{0.7cm}

\begin{center}
Jean-Emile B{\sc ourgine}\footnote
            {
e-mail address : 
jebourgine@apctp.org}\\
      
\vspace{0.7cm}                    
{\it Asia Pacific Center for Theoretical Physics (APCTP)
}\\
{\it Pohang, Gyeongbuk 790-784, Republic of Korea},

\end{center}

\vspace{0.7cm}

\begin{abstract}
\noindent
The Spherical Hecke central (SHc) algebra has been shown to act on the Nekrasov instanton partition functions of $\mathcal{N}=2$ gauge theories. Its presence accounts for both integrability and AGT correspondence. On the other hand, a specific limit of the Omega background, introduced by Nekrasov and Shatashvili (NS), leads to the appearance of TBA and Bethe like equations. To unify these two points of view, we study the NS limit of the SHc algebra. We provide an expression of the instanton partition function in terms of Bethe roots, and define a set of operators that generates infinitesimal variations of the roots. These operators obey the commutation relations defining the SHc algebra at first order in the equivariant parameter $\e_2$. Furthermore, their action on the bifundamental contributions reproduces the Kanno-Matsuo-Zhang transformation. We also discuss the connections with the Mayer cluster expansion approach that leads to TBA-like equations.

\vspace{0.5cm}
\end{abstract}

\vfill

\end{titlepage}
\vfil\eject

\setcounter{footnote}{0}

\section{Introduction}
In many respects, degenerate affine Hecke algebras seem to be the key behind the fascinating structure of instanton partition functions of 4d $\mathcal{N}=2$ gauge theories. The Spherical Hecke central (SHc) algebra constructed in \cite{Schiffmann2012} is a limit of a symmetrized double degenerate Hecke algebra (DDAHA). It acts on instanton partition functions as the Kanno-Matsuo-Zhang (KMZ) transformation of the Young diagrams summand, also called \textit{bifundamental contribution} \cite{Kanno2013}. SHc representations contain $W_N$ subalgebras which is the main reason behind the AGT correspondence \cite{Alday2009,Wyllard2009}. And indeed, the various proofs of AGT correspondence exploit this underlying algebraic structure. For instance, a set of (generalized) Jack polynomials have been used in the proofs \cite{Mironov2010d,Zhang2011,Mironov2012c,Morozov2013,Mironov2013} based on the free field representation of CFT conformal blocks \cite{Dotsenko1984,Dotsenko1984a}. Jack polynomials are known to be the eigenstates of the Calogero-Moser Hamiltonian which is one of the SHc generators in the polynomial representation relevant to the formal case of gauge groups $SU(\Nc)$ with $\Nc=1$ \cite{Schiffmann2012}. The SHc algebra has a Hopf algebra structure, and it is possible to employ the comultiplication to define a representation of the generators relevant to the case of a higher number of colors ($\Nc>1$). This leads to define generalized Jack polynomials which diagonalize the Calogero-Moser Hamiltonian acting in a tensor space \cite{Morozov2013,Smirnov2014}. The first proof of AGT correspondence is also based indirectly on SHc. It uses a specific basis of the CFT Fock space formed by the AFLT states. These states were defined such that the decomposition of conformal blocks coincides with the expression of Nekrasov partition functions \cite{Fateev2009,Alba2010,Fateev2011,Kanno2011}. They were later identified with the generalized Jack polynomials \cite{Belavin2011a,Smirnov2014} using the free field construction presented in \cite{Awata1994,Awata1995}.

DDAHA, from which SHc is constructed, is formed by Dunkl operators, and contains an infinite number of commuting integrals of motion. It has a deep connection with the quantum inverse scattering method \cite{Bernard1993,Pasquier1994,Cherednik2005}, which may lead to a better understanding of the instanton R-matrix proposed by Smirnov \cite{Smirnov2013} and built upon the stable map introduced in \cite{Maulik2012}.

In this paper, we are interested in yet another aspect of SHc. In \cite{Nekrasov2009}, Nekrasov and Shatashvili (NS) have investigated a limit of the $\O$-background in which one of the equivariant parameters tends to zero. They have proposed to express the gauge theory free energy in this limit using the solution of a non-linear integral equation (NLIE) reminiscent of those derived from the Thermodynamical Bethe Ansatz \cite{Zamolodchikov1990a,Destri1994}. This proposal was confirmed in \cite{Meneghelli2013,Bourgine2014}, exploiting the Mayer cluster expansion technique to perform the subtle limit. In a different study of the NS limit, Bethe-like equations and TQ-relations were also obtained \cite{Poghossian2010,Fucito2011,Fucito2012,Nekrasov2013}. In both cases, the associated integrable model bare some similarities with the $sl(N)$ XXX spin chain, and the system of bosons with delta interaction studied by C. N. Yang \cite{Gaudin1983}.\footnote{Both systems share many common properties. For instance, Yang-Yang functionals have the same quadratic part, and differ only by the potential term. Coordinate wave functions are also very close. The gauge theory seems to appeal to what these two systems have in common.} The latter also describes the solution of the quantum non-linear Schr\"odinger equation in a sector of fixed number of particles, and will be referred shortly as \textit{qNLS}. Both integrable models are related to Hecke algebras, although in a different way. XXX spin chains (with open boundaries) are known to possess a Yangian symmetry, which is a representation of the degenerate affine Hecke algebra \cite{Guay2005}. On the other hand, the qNLS Hamiltonian can be diagonalized by a technique involving Dunkl operators. In the case of periodic boundary conditions, these operators form with the affine Weyl group a DDAHA \cite{Emsiz2005}. In view of these results, it seems necessary to understand the fate of the SHc algebra in the NS limit, and possibly relate it to the algebraic structure of XXX and qNLS integrable models. This paper reports on the first step in this direction.

We focus on a $A_2$ quiver gauge theory with $U(\Nc)\times U(\Nc)$ gauge group, $\Nc$ fundamental flavors at each node, and a bifundamental matter field. This theory is asymptotically conformal. We regard more precisely the bifundamental contribution to the Nekrasov instanton partition function. This quantity depends on two sets of $\Nc$ Young tableaux, and is a building block for the instanton partition function of more general quivers. The KMZ transformation represents the action of SHc generators as a variation of the number of boxes in the Young diagrams. In order to proceed to the NS limit, the action of these generators must be re-written in a suitable manner. This is done in the second section. In the third section, we recall the derivation of the Bethe-like equations from the invariance of the Young diagrams summands under a variation of boxes. It leads to identify the Bethe roots with the instanton positions of the boxes on top of each column. This identification further provides an expression of the bifundamental contribution. We then construct a set of operators upon infinitesimal variations of the Bethe roots, and show that they obey the SHc commutation relations at first order in $\e_2$. We also recover the KMZ transformation in this limit. Finally, in a fourth section, we explain the connection with the outcome of the Mayer cluster expansion, and TBA-like NLIE. Technical details are gathered in the appendices.

\section{Spherical Hecke central algebra}
The main properties of the SHc algebra can be found in \cite{Kanno2013} which is the starting point for our rewriting process. In order to render the $\e_2$ factors explicit, we will avoid the notations $\b$ and $\xi_\text{KMZ}=1-\b$, and use instead directly the $\O$-background equivariant parameters $\e_1$ and $\e_2$, together with the shortcut notation $\e_+=\e_1+\e_2$. These parameters are related to the previous quantities through $\b=-\e_1/\e_2$ and $ \xi_\text{KMZ}=\e_+/\e_2$. Let us emphasize that no limit is taken in this section, and expressions are exact in $\e_1$ and $\e_2$.

\subsection{Rewriting the SHc algebra}
\subsubsection{Partition function}
The building block of quiver partition functions is the so-called \textit{bifundamental} contribution.\footnote{Strictly speaking, this quantity also contains a contribution from the vector multiplets coupled to the bifundamental multiplet in its denominator. These normalization factors simplify the expression of the KMZ transformation.} It is associated to an arrow of the quiver between two nodes $a\to b$. It depends on the information at each node $a$, encoded in an object $Y_a$, and a mass $m_{ab}$ of bifundamental matter fields. Here we focus on the $A_2$ quiver, with two nodes $a=1,2$, and a single fundamental mass $m$ for which we drop the index $12$. To each node $a$ corresponds a gauge group $U(N_c^{(a)})$, and a vector of Coulomb branch vevs $a_l^{(a)}$ with $l=1\cdots \Nc^{(a)}$ the color index. We further associate to a node $a$ the object denoted $Y_a$ that consists in a set of $\Nc^{(a)}$ Young diagrams $Y_a^{(l)}$, together with the vector $a_l^{(a)}$. Young diagrams $Y_a^{(l)}$ are made of $n_l^{(a)}$ columns $\l_i^{(a,l)}$ with $\l_{i+1}^{(a,l)}\leq\l_i^{(a,l)}$. The dual partition consists of the sequence of integers $\tl_i^{(a,l)}=\sharp\{j\diagup i\leq\l_j^{(a,l)}\}$ for $i=1\cdots \l_1^{(a,l)}$. Indices $i$ of $\l_i^{(a,l)}$ and $\tl_i^{(a,l)}$ can be extended to infinity, setting the additional quantities to be zero. For simplicity, we assume that both nodes have the same gauge group, i.e. $\Nc^{(1)}=\Nc^{(2)}=\Nc$.

The bifundamental contribution can be expressed in terms of the variables $t_{l,i}^{(a)}$ and the dual ones $\tt_{l,i}^{(a)}$ associated to $Y_a$, and defined as
\begin{equation}\label{def_tli}
t_{l,i}^{(a)}=a_l^{(a)}+\e_1(i-1)+\e_2\l_i^{(a,l)},\quad \tt_{l,j}^{(a)}=a_l^{(a)}+\e_1\tl_j^{(a,l)}+\e_2(j-1),
\end{equation}
where $i=1\cdots n_l^{(a)}$ spans the columns, and $j=1\cdots \l_0^{(a,l)}$ the rows of the diagram $Y_a^{(l)}$. It reads
\begin{equation}\label{CZ_YY}
\CZ[Y_1,Y_2]=\dfrac{\prod_{l,(i,j)\in Y_1}\prod_{l'=1}^{\Nc}(\tt_{l,j}^{(1)}-t_{l',i}^{(2)}-\mu+\e_2)\prod_{l,(i,j)\in Y_2}\prod_{l'=1}^{\Nc}(\tt_{l,j}^{(2)}-t_{l',i}^{(1)}+\mu-\e_1)}{\prod_{a=1,2}\left[\prod_{l,(i,j)\in Y_a}\prod_{l'=1}^{\Nc}(\tt_{l,j}^{(a)}-t_{l',i}^{(a)}+\e_2)(\tt_{l,j}^{(a)}-t_{l',i}^{(a)}-\e_1)\right]^{1/2}},
\end{equation}
with $|Y_a|$ the total number of boxes of Young diagrams. This expression has been obtained from the equation (6) of \cite{Kanno2013} under a rescaling of Coulomb branch vevs and bifundamental mass: $a_l^{(a)}=-\e_2a_{p,\text{KMZ}}$ and $\mu=-\e_2\mu_\text{KMZ}$.\footnote{This rescaling is necessary in order to have a non-trivial dependence. Correspondingly, the study of semiclassical Liouville correlators with heavy operators is more involved than the case of light ones.} The variable $\mu$ corresponds to a shifted bifundamental mass, it is related to $m$ through $\mu=\e_+/2-m$. In the following we mostly focus on a single object $Y_{a=1}$ and, to alleviate the notations, the label $a$ will be dropped when no ambiguity arise.

\subsubsection{Instanton positions and $\L$-factors}
The SHc algebra found in \cite{Kanno2013} is acting on states $|Y>$ characterized by an object $Y$ (we dropped the node index $a$). To each box $x\in Y$ is associated a triplet of indices $l,i,j$ such that $(i,j)\in Y^{(l)}$ is a box in the $l$th Young diagram. In our reformulation of the SHc algebra, a prominent role is played by the instanton position which is a map that associates to each box $x\in Y$ the complex number 
\begin{equation}\label{map_phix}
\phi_x=a_l+(i-1)\e_1+(j-1)\e_2.
\end{equation}
Thus, the variables $t_{l,i}$ defined in \ref{def_tli} correspond to the instanton positions of the box locations that lay on top of each column $\l_i^{(l)}$. Similarly, $\tt_{l,j}$ is the instanton position of a box located directly on the right of each row.

We denote $A(Y)$ (resp. $R(Y)$) the set of boxes that can be added to (resp. removed from) $Y$. The set $A(Y)$ is a subset of the locations with positions $t_{l,i}$, and also of the locations with positions $\tt_{l,j}$. Likewise, the set $R(Y)$ is a subset of the locations with positions $t_{l,i}-\e_2$, and $\tt_{l,j}-\e_1$:
\begin{align}
\begin{split}
&A(Y)\subset\{x\in Y\diagup\phi_x=t_{l,i}\},\quad A(Y)\subset\{x\in Y\diagup\phi_x=\tt_{l,j}\},\\
&R(Y)\subset\{x\in Y\diagup\phi_x=t_{l,i}-\e_2\},\quad R(Y)\subset\{x\in Y\diagup\phi_x=\tt_{l,j}-\e_1\}.
\end{split}
\end{align}
Note that for each Young diagram the number of boxes one can add is always one plus the number we can remove: $\sharp A(Y)=\sharp R(Y)+\Nc$ where $\Nc$ is the number of Young diagrams in $Y$.

In the original paper \cite{Kanno2013}, the KMZ transformation is expressed using a decomposition of Young diagrams into elementary rectangles. This rectangle decomposition is useful to characterize the coordinates of boxes in the sets $A(Y)$ and $R(Y)$. However, it turns out that this information is not essential to the formulation of the KMZ transformation. It is sufficient to use the instanton positions $\phi_x$ for $x\in A(Y)$ or $x\in R(Y)$ without specifying the coordinates of these boxes in the Young diagrams. Omitting this information results in simpler expressions for the SHc generators and the KMZ transformation. The transition between the notations of the KMZ paper \cite{Kanno2013} and ours is explained in the appendix \refOld{App0}. It may be summarized by the formulas \ref{transition_1} and \ref{transition_2}.

One of the main simplifications in our formalism concerns the factors $\L_l^{(k,\pm)}$ given by equ (12) and (13) in \cite{Kanno2013} that can be merged into a single quantity,
\begin{equation}
\L_x(Y)^2=\prod_{\superp{y\in A(Y)}{y\neq x}}\dfrac{\phi_x-\phi_y+\e_+}{\phi_x-\phi_y}\prod_{\superp{y\in R(Y)}{y\neq x}}\dfrac{\phi_x-\phi_y-\e_+}{\phi_x-\phi_y}.
\end{equation}
Note that the term $y=x$ has to be removed from the left product if $x\in A(Y)$ and the right one for $x\in R(Y)$. The expression remains valid if $x\notin A(Y)\cup R(Y)$ with both products complete. In the limit $\e_+=0$ (or $\b=1$), we trivially have $\L_x(Y)=1$. The factors $\L_x(Y)$ play a major role in the definition of the SHc algebra, and it is important to study their properties. They correspond to the residues at the poles of the function
\begin{equation}
\L(z)^2=\prod_{x\in A(Y)}\dfrac{z-\phi_x+\e_+}{z-\phi_x}\prod_{x\in R(Y)}\dfrac{z-\phi_x-\e_+}{z-\phi_x},
\end{equation}
which can be decomposed as
\begin{equation}\label{prop_L}
\L(z)^2=1+\e_+\sum_{x\in A(Y)}\dfrac{\L_x(Y)^2}{z-\phi_x}-\e_+\sum_{x\in R(Y)}\dfrac{\L_x(Y)^2}{z-\phi_x}.
\end{equation}
An infinite tower of identities can be obtained through an expansion at $z\to\infty$. The first identities are given in the appendix \refOld{App0}.

Operators of the SHc algebra act on the states $|Y>$ by adding or removing boxes. We denote $Y+x$ and $Y-x$ the object $Y$ with a box $x\in A(Y)$ (resp $x\in R(Y)$) added to (removed from) the set of Young diagrams. The behavior of the coefficients $\L_x(Y)$ under such a procedure is remarkably simple:
\begin{equation}\label{shift_L}
\L_y(Y-x)^2=r(\phi_x-\phi_y)\L_y(Y)^2,\quad \L_y(Y+x)^2=r(\phi_y-\phi_x)\L_y(Y)^2,
\end{equation}
where we introduced the function
\begin{equation}\label{def_rz}
r(z)=\dfrac{(z+\e_1)(z+\e_2)(z-\e_+)}{(z-\e_1)(z-\e_2)(z+\e_+)},\quad r(-z)=1/r(z).
\end{equation}
It is important to note that the quantity $r(\phi_x-\phi_y)$ is independent of the Young diagrams. Let $x\in R(Y)$, the first formula in \ref{shift_L} is valid if $y\in A(Y-x)\cup A(Y)$ or $y\in R(Y-x)\cup R(Y)$. It breaks down if $y\in R(Y-x)$ but $y\notin R(Y)$ which corresponds to the poles of $r(\phi_x-\phi_y)$ for $\phi_x-\phi_y=\e_1$ or $\e_2$ (boxes under $x$ or on the left). Extending the definition of $\L_y(Y)^2$ to $x\notin Y$, $r(z)$ in \ref{shift_L} has to be replaced by its residue divided by $\e_+$. When $y\in A(Y-x)$ but $y\notin A(Y)$, i.e. $y=x$, we have to replace $r(0)=-1$ by one. The same comment is valid for $x\in A(Y)$: if $y\in A(Y+x)$ but $y\notin A(Y)$, it implies that $\phi_y-\phi_x=\e_1$ or $\e_2$ (boxes above $x$ or on its right) and the pole of $r(z)$ has to be replaced by the residue divided by $\e_+$. For $y=x\in R(Y+x)$, we have $\L_x(Y+x)^2=\L_x(Y)^2$.

\subsubsection{Generators}
We are now ready to introduce the generators $D_{n,m}$ of the SHc algebra. The index $n$ runs over all integers and is referred as the \textit{degree}, whereas the second index $m$ is a positive integer called the \textit{order}. The action of these generators on the states $|Y>$ is known in a closed form only for the operators of degrees $0$ and $\pm1$. For those operators we introduce the generating series
\begin{equation}\label{def_D}
D_{\pm1}(z)=\sum_{n=0}^\infty{(z+\e_+)^{-n-1}\e_2^n D_{\pm1,n}},\quad D_0(z)=\sum_{n=0}^\infty{(z+\e_+)^{-n-1}\e_2^n D_{0,n+1}}.
\end{equation}
The construction of these operators from the degenerate affine Hecke algebra is done in \cite{Schiffmann2012}. For $N_c=1$, the generators $D_{0,n}$ in the polynomial representation are symmetric polynomials of the Cherednik(-Dunkl) operators for the Calogero-Moser model. In particular, $D_{0,2}$ coincide with the Hamiltonian of this integrable system, and the states $|Y>$ can be identified with the eigenbasis of Jack polynomials. The operators $D_{\pm1,n}$ are built by exploiting the Pieiri formula satisfied by Jack polynomials \cite{Stanley1989}. At $N_c>1$, it is possible to exploit the Hopf algebra structure of SHc to define a tensor representation for $D_{0,n}$ \cite{Schiffmann2012}. Then, the states $|Y>$ should be identified with the generalized Jack polynomials which diagonalize the comultiplication of $D_{0,2}$ \cite{Morozov2013,Smirnov2014,Belavin2011a}.

The action of the generators on states $|Y>$ is defined by the formulas (38)-(40) of \cite{Kanno2013}, which can be rewritten as
\begin{equation}\label{act_D}
D_{\pm1}(z)|Y>=\sum_{x\in A/R(Y)}\dfrac{\L_x(Y)}{z-\phi_x}|Y\pm x>,\quad D_0(z)|Y>=\sum_{x\in Y}\dfrac1{z-\phi_x}|Y>.
\end{equation}
The commutation relations (22)-(23) of \cite{Kanno2013} can be summed up to produce
\begin{equation}\label{comm_SHc}
[D_0(z),D_{\pm1}(w)]=\pm\dfrac{D_{\pm1}(w)-D_{\pm1}(z)}{z-w},\quad [D_{-1}(z),D_1(w)]=\dfrac{E(w)-E(z)}{z-w},
\end{equation}
where we defined another generating series:
\begin{equation}\label{def_E}
E(z)=\sum_{n=0}^\infty{(z+\e_+)^{-n-1}\e_2^n E_n}.
\end{equation}

The specific properties of the SHc algebra comes from the expressions of the generators $E_n$ as a function of degree zero operators. They are encoded in the identity
\begin{equation}\label{rel_E_D0}
1+\e_+E(z)=\exp\left(\sum_{n\geq0}(-1)^{n+1}c_n\pi_n(\e_2/(z+\e_+))\right)\exp\left(\sum_{n\geq0}D_{0,n+1}\o_n(\e_2/(z+\e_+))\right),
\end{equation}
with the functions
\begin{align}
\begin{split}
&\pi_n(s)=s^nG_n(1+s\e_+/\e_2),\quad \o_n(s)=\sum_{q=\e_2,\e_1,-\e_+}s^n\left(G_n(1-qs/\e_2)-G_n(1+qs/\e_2)\right),\\
&G_0(s)=-\log(s),\quad G_n(s)=(s^{-n}-1)/n,\quad (n\geq1).
\end{split}
\end{align}
The central charges $c_n$ are the Miwa transformed of the Coulomb branch vevs,
\begin{equation}
\e_2^nc_n=(-1)^n\sum_{l=1}^{\Nc}(a_l+\e_+)^n.
\end{equation}
It is possible to deduce the action of $E(z)$ on states $|Y>$ from \ref{rel_E_D0} and the action \ref{act_D} of $D_0(z)$:\footnote{We have used the formula
\begin{equation}\label{form_Gn}
\log\left(\dfrac{1+(a+b)s}{1+as}\right)=\sum_{n\geq0}(-1)^{n+1}a^ns^nG_n(1+bs).
\end{equation}}
\begin{equation}\label{vp_E0}
\left(1+\e_+E(z)\right)|Y>=\prod_{l=1}^{\Nc}\left(1+\dfrac{\e_+}{z-a_l}\right)\prod_{x\in Y}r(z-\phi_x)|Y>,
\end{equation}
with the function $r(z)$ defined in \ref{def_rz}. In the RHS, the product over $r(z-\phi_x)$ and Coulomb branch vevs reduces to the function $\L(z)^2$ as a result of the identity\footnote{This identity is easy to derive for Young diagrams that are rectangles. It proceeds from the numerous cancellations between numerators and denominators in the LHS product. For an arbitrary Young diagram, the equality is obtained by decomposition into elementary rectangles.}
\begin{equation}\label{rel_r_L}
\prod_{x\in Y}r(z-\phi_x)=\L(z)^2\prod_{l=1}^{\Nc}\dfrac{z-a_l}{z-a_l+\e_+}.
\end{equation}

\begin{figure}[!t]
\centering
\includegraphics[width=9cm]{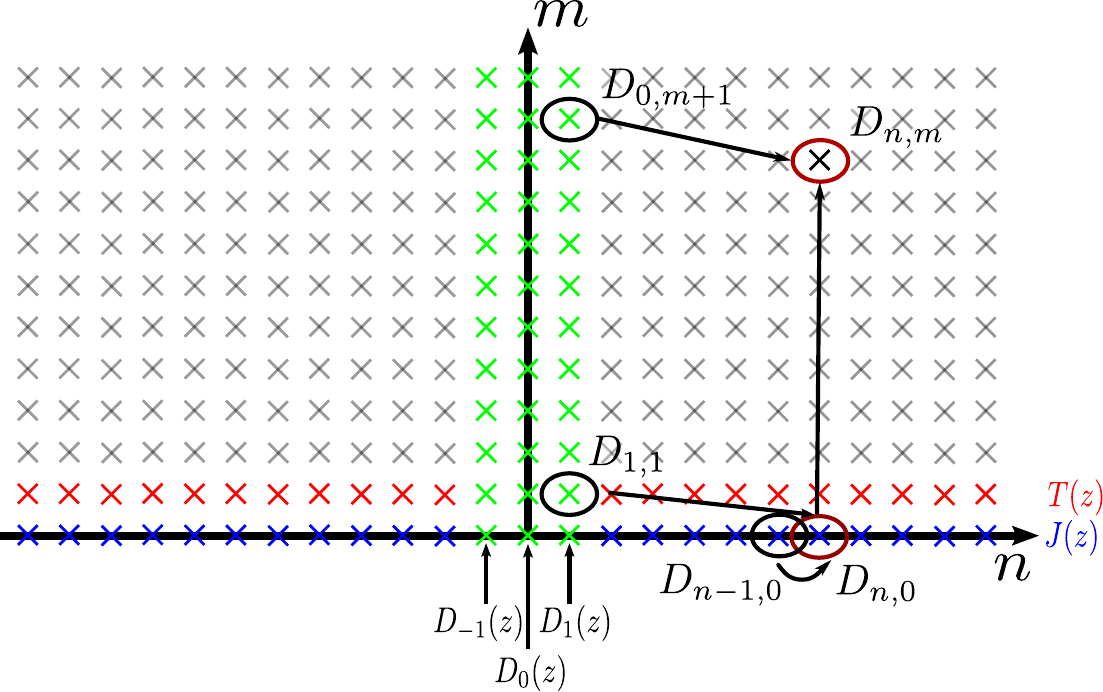}
\caption{Location of the operators $D_{n,m}$ in the $\mathbb{Z}\times\mathbb{Z}^+$ half-plane. In green, the generators forming $D_0(z)$ and $D_{\pm1}(z)$ for which the action on the states $|Y>$ is known. In blue and red, the modes of the $U(1)$ current and stress-energy tensor respectively. Black arrows represent the commutation relations \ref{def_hg}.}
\label{fig_d2}
\end{figure}

Operators of higher degree are constructed using the following commutation relations $(n\geq0, m>0)$:
\begin{equation}\label{def_hg}
D_{\pm(n+1),0}=\pm\dfrac1n[D_{\pm1,1},D_{\pm n,0}],\quad D_{\pm n,m}=\pm[D_{0,m+1},D_{\pm n,0}].
\end{equation}
From these commutation relations, the generators of degree zero and $\pm1$ engenders the whole algebra. In particular, the $U(1)$ and Virasoro subalgebras are generated by the modes of the current $J_n$ and stress-energy tensor $L_n$ defined with generators of \textbf{order} zero and one:
\begin{equation}
J_{\pm n}=(-\sqrt{-\e_1/\e_2})^{-n}D_{\mp n,0},\quad L_{\pm n}=\dfrac1n(-\sqrt{-\e_1/\e_2})^{-n}D_{\mp n,1}+\hf(1-n)\Nc\dfrac{\e_+}{\e_2}J_{\pm n}.
\end{equation}
A summary of the different relations between the SHc generators can be found in figure \refOld{fig_d2}. The generators of Virasoro and $U(1)$ subalgebras are orthogonal to the operators composing the series $D_0(z)$ and $D_{\pm1}(z)$. The standard current and Virasoro modes commutation relations can be recovered from \ref{comm_SHc} and \ref{def_hg}. However, this nested calculation rapidly becomes impractical.

\subsection{KMZ transformation}
The KMZ transformation is generated by the operators $\d_{\pm1,n}$ in \cite{Kanno2013}. Here, we use instead the generating series
\begin{equation}
\d_{\pm1,z}=\sum_{n=0}^\infty (z+\e_+)^{-n-1}(-\e_2)^n\d_{\pm1,n}.
\end{equation}
In our notations, they act on the bifundamental contributions \ref{CZ_YY} as
\begin{align}
\begin{split}\label{KMZ_CZ}
\d_{-1,z}\CZ[Y_1,Y_2]&=\sum_{x\in A(Y_1)}\dfrac{\L_x(Y_1)}{z-\phi_x}\CZ[Y_1+x,Y_2]-\sum_{x\in R(Y_2)}\dfrac{\L_x(Y_2)}{z-\mu-\phi_x}\CZ[Y_1,Y_2-x].\\
\d_{+1,z}\CZ[Y_1,Y_2]&=-\sum_{x\in R(Y_1)}\dfrac{\L_x(Y_1)}{z-\phi_x}\CZ[Y_1-x,Y_2]+\sum_{x\in A(Y_2)}\dfrac{\L_x(Y_2)}{z+\e_+-\mu-\phi_x}\CZ[Y_1,Y_2+x].
\end{split}
\end{align}
The operator $\d_{-1,z}$ combines an action of $D_{+1}(z)$ on the state $|Y_1>$ (first term) and an action of $D_{-1}(z-\mu)$ on $|Y_2>$ (second term). In $\d_{+1,z}$, the actions of $D_{\pm1}(z)$ are exchanged: $D_{-1}(z)$ now acts on $|Y_1>$ and $D_{+1}(z+\e_+-\mu)$ on $|Y_2>$. It is possible to work at fixed box $x$ using a contour integral circling $\phi_x$. For $x\in A/R(Y_1)$, we have
\begin{equation}
\oint_{\phi_x}{\dfrac{dz}{2i\pi}\ \d_{\mp1,z}\CZ[Y_1,Y_2]}=\pm\L_x(Y_1)\CZ[Y_1\pm x,Y_2].
\end{equation}

The covariance of the bifundamental contribution $\CZ[Y_1,Y_2]$ under the KMZ transformation involves the functions
\begin{align}
\begin{split}\label{def_U}
U_{-1,z}&=\dfrac{\prod_{x\in R(Y_1)}(z-\e_+-\phi_x)\prod_{x\in A(Y_2)}(z-\mu+\e_+-\phi_x)}{\prod_{x\in A(Y_1)}(z-\phi_x)\prod_{x\in R(Y_2)}(z-\mu-\phi_x)}-1,\\
U_{+1,z}&=\dfrac{\prod_{x\in A(Y_1)}(z+\e_+-\phi_x)\prod_{x\in R(Y_2)}(z-\mu-\phi_x)}{\prod_{x\in R(Y_1)}(z-\phi_x)\prod_{x\in A(Y_2)}(z-\mu+\e_+-\phi_x)}-1.
\end{split}
\end{align}
Note that the minus one in the RHS subtracts the pole at $z=\infty$. Under the KMZ transformation, $\CZ[Y_1,Y_2]$ behaves as
\begin{equation}\label{KMZ}
\sqrt{-\e_1\e_2}\d_{\pm1,z}\CZ[Y_1,Y_2]=U_{\pm1,z}\CZ[Y_1,Y_2].
\end{equation}
At fixed box $x\in A(Y_1)$ this relation writes
\begin{equation}\label{CZ_YY_shift1}
\sqrt{-\e_1\e_2}\L_x(Y_1)\CZ[Y_1+x,Y_2]=\dfrac{\prod_{y\in R(Y_1)}(\phi_x-\phi_y-\e_+)\prod_{y\in A(Y_2)}(\phi_x-\phi_y-\mu+\e_+)}{\prod_{\superp{y\in A(Y_1)}{y\neq x}}(\phi_x-\phi_y)\prod_{y\in R(Y_2)}(\phi_x-\phi_y-\mu)}\CZ[Y_1,Y_2].
\end{equation}
A similar relation can be obtained for $x\in A(Y_2)$ by exchanging $Y_1\leftrightarrow Y_2$, and replacing $\mu\to \e_+-\mu$.

\section{SHc in the NS limit}
We now recall the derivation of the Bethe equations in the NS limit along the lines of \cite{Chen2012,Nekrasov2013}. The derivation exploits the covariance of summands of the partition function under variation of the number of boxes in Young diagrams. We argue that the main contribution to the partition function comes from Young diagrams with columns of infinite height, and multiplicity one. The profile of these diagrams is fully encoded in a set of Bethe roots which will be used to express the partition function.

\subsection{Derivation of the Bethe equations}
In \cite{Nekrasov2003a}, Nekrasov and Okounkov have exposed a procedure to compute the instanton partition function in the Seiberg-Witten limit $\e_1,\e_2\to0$. In this limit, Young diagrams become infinitely large, and are described by a continuous profile minimizing an effective action. This method was later extended to the NS limit $\e_2\to0$ in \cite{Poghossian2010,Fucito2011,Fucito2012,Ferrari2012a}.\footnote{It was also employed in \cite{Piatek2014a,Piatek2014}.} It relies on the idea that Young diagrams summations are dominated by a set of Young diagrams with a specific profile. This profile is an extremum, which implies that small deformations of the profile, typically by adding or removing some boxes, vanish at first order. And indeed the requirement of invariance under adding or removing boxes leads to Bethe-like equations that characterizes the NS limit \cite{Chen2012,Nekrasov2013}. We review this method here.

The instanton partition function of the $A_2$ quiver reads
\begin{equation}\label{decomp_CZ}
\CZ_{A_2}[M_1,M_2]=\sum_{Y_1,Y_2}q_1^{|Y_1|}q_2^{|Y_2|}\CZ[M_1,Y_1]\CZ[Y_1,Y_2]\CZ[Y_2,M_2],
\end{equation}
where the sum is over all the possible realizations of $Y_{a=1,2}$ for a fixed set of Coulomb branch vevs. The objects $M_a$ encode the fundamental flavors content of the theory. They consist of $N_f^{(a)}$ empty Young diagrams, and a vector $m_l^{(a)}$ with $l=1\cdots N_f^{(a)}$ corresponding to the masses of the $N_f^{(a)}$ flavors in the fundamental representation. They are such that $R(M_a)=\varnothing$ and $A(M_a)=\{(l,1,1),l=1\cdots N_f^{(a)}\}$. To be able to use the transformation properties of the previous subsection, we need to require $N_f^{(1)}=N_f^{(2)}=\Nc$. In addition, the bifundamental mass $\mu$ is equal to zero for both $\CZ[M_1,Y_1]$ and $\CZ[Y_2,M_2]$.

The variation of the summand under addition of a box $x\in A(Y_1)$ can be computed using the KMZ transformation \ref{CZ_YY_shift1} of bifundamental contributions:
\begin{align}
\begin{split}
\label{var_CZ_YY}
&\dfrac{q_1^{|Y_1+x|}\CZ[M_1,Y_1+x]\CZ[Y_1+x,Y_2]}{q_1^{|Y_1|}\CZ[M_1,Y_1]\CZ[Y_1,Y_2]}\\
=&-\dfrac{q_1}{\e_1\e_2}m_1(\phi_x)\dfrac{\prod_{y\in R(Y_1)}(\phi_x-\phi_y)(\phi_x-\phi_y-\e_+)}{\prod_{\superp{y\in A(Y_1)}{y\neq x}}(\phi_x-\phi_y)(\phi_x-\phi_y+\e_+)}\dfrac{\prod_{y\in A(Y_2)}(\phi_x-\phi_y-\mu+\e_+)}{\prod_{y\in R(Y_2)}(\phi_x-\phi_y-\mu)}.
\end{split}
\end{align}
The mass polynomial $m_a(x)$ can be found in \ref{pot_quiver} below, it is monic, with zeros at the value of the fundamental masses $x=m_l^{(a)}$ for $l=1\cdots N_f^{(a)}$. To recover the Bethe equations from the condition
\begin{equation}
\dfrac{q_1^{|Y_1+x|}\CZ[M_1,Y_1+x]\CZ[Y_1+x,Y_2]}{q_1^{|Y_1|}\CZ[M_1,Y_1]\CZ[Y_1,Y_2]}=1,
\end{equation}
we assume that the configurations of Young diagrams that contribute in the NS limit are such that all columns have a different height. As a consequence, a box can be added or removed from any column. Furthermore, the columns heights are sent to infinity, such that the products $\e_2\l_i^{(l,a)}$ remain finite. We then identify the Bethe roots with the instanton positions of the boxes on top of each column. These also coincide with the position of boxes that can be added and removed, up to a negligible shift in $\e_2$:
\begin{equation}\label{id_sets}
u_r=t_{l,i}=\phi_x,\quad x\in \tilde{A}(Y)\text{ or }R(Y),\qquad\text{in the NS limit}.
\end{equation}
Such an identification was actually already made in \cite{Nekrasov2013,Bourgine2014}. The assumption on the shape of Young diagrams will be justified by the consistency observed with the results of the Mayer cluster expansion. There is however a simple heuristic explanation: to keep $t_{l,i}$ finite as $\e_2\to0$, we effectively rescaled the Young diagrams columns by $1/\e_2$ which increases the disparities between them.\footnote{Columns with equal heights produce Bethe roots spaced of $\e_1$, which should correspond to strings solutions of the Bethe equations. A priori, there are no such solutions for the Bethe equations \ref{Bethe_1}, but it could be interesting to perform a deeper analysis of some degenerate situations.}

There is however a small additional subtlety, since there is always one more box that can be added to a Young diagram than that can be removed. For each $Y_a$, these $N_c$ extra boxes lie on the right of the diagrams, at positions $\xi_l^{(a)}=a_l^{(a)}+n_l^{(a)}\e_1$. In \ref{id_sets}, we have specified the set $\tilde{A}(Y)=A(Y)\setminus\{x\in Y\diagup \phi_x=\xi_l\}$, excluding the extra boxes at position $\xi_l$. If we are considering Young diagrams with infinitely many columns, implying that the number of Bethe roots is also infinite, these extra boxes can be neglected and $A(Y)\simeq \tilde{A}(Y)$. Here, we will keep the number of columns finite, and $n_l^{(a)}$ will act as a cut-off, as in \cite{Poghossian2010,Fucito2011,Bourgine2012a}.
%

To each node of the quiver is associated a set of Bethe roots. For the $A_2$ quiver, we denote the two sets of Bethe roots $u_r$ and $v_r$ for the nodes $1$ and $2$ respectively, with $r=1\cdots N_B^{(a)}$. From this identification, we find for $\phi_x\to u_r$ that the RHS of \ref{var_CZ_YY} becomes\footnote{The factor $-\e_1\e_2$ comes from the box $y\in R(Y_1)$ just below $x$ which is such that $\phi_y=\phi_x-\e_2$.}
\begin{align}
\begin{split}
\label{Bethe_1}
&q_1m_1(u_r)\xi_1(u_r)\prod_{\superp{s=1}{s\neq r}}^{\NB^{(1)}}\dfrac{u_r-u_s-\e_1}{u_r-u_s+\e_1}\prod_{s=1}^{\NB^{(2)}}\dfrac{u_r-v_s+m+\e_1/2}{u_r-v_s+m-\e_1/2}=1,\\
&q_2m_2(v_r)\xi_2(v_r)\prod_{\superp{s=1}{s\neq r}}^{\NB^{(2)}}\dfrac{v_r-v_s-\e_1}{v_r-v_s+\e_1}\prod_{s=1}^{\NB^{(1)}}\dfrac{v_r-u_s-m+\e_1/2}{v_r-u_s-m-\e_1/2}=1.
\end{split}
\end{align}
The second equation is obtained from the variation of the profile of $Y_2$, it is identical to the first one with $u_r$ and $v_r$ exchanged, and $m\to -m$. These two sets of equations form the so-called Bethe-like equations of the $A_2$ quiver. The factors $\xi_a(x)$, given by ($m_{21}=-m_{12}=-m$)
\begin{equation}
\xi_a(x)=\prod_{l=1}^{\Nc}\dfrac{\prod_{b\neq a}(x-\xi_l^{(b)}+m_{ab}+\e_1/2)}{(x-\xi_l^{(a)})(x+\e_1-\xi_l^{(a)})},
\end{equation}
disappear for infinitely large Young diagrams.

The Young diagrams with infinite profiles are the natural states on which the SHc generators act in the NS limit. Let us consider a regular object $Y$ (with a finite number of boxes), and pick a box $x=(l,i,j)\in A(Y)\cup R(Y)$ with coordinates $(i,j)\in Y^{(l)}$. It is shown in the appendix \refOld{AppB0} that $\L_x(Y)=O(\sqrt{\e_2})$ unless $i=1$ which gives $\L_x(Y)=O(1)$. Thus, the summation of boxes $x\in A/R(Y)$ defining the action \ref{act_D} of $D_{\pm1}(z)$ on the states $|Y>$ can be replaced by a much simpler summation over the number of colors. The argument breaks down for Young tableaux of infinite profile, for which the difference of height between two neighboring columns times $\e_2$ becomes macroscopic. Then, $\L_x(Y)$ is of order $O(1/\sqrt{\e_2})$ for all boxes in $A(Y)\cup R(Y)$, and the action of $D_{\pm1}(z)$ on the corresponding states involves a non-trivial summation over the whole set of Bethe roots.

To summarize, we have found that for any test function $f$, we have for the NS limit of the object $Y$:
\begin{equation}\label{dico_ur}
\sum_{x\in \tilde{A}/R(Y)}{f(\phi_x)}\to\sum_{r=1}^{N_B}{f(u_r)},\quad \sum_{x\in A(Y)}{f(\phi_x)}\to\sum_{r=1}^{N_B}{f(u_r)}+\sum_{l=1}^{N_c}f(\xi_l).
\end{equation}
In particular, the function $\L(z)^2$ becomes:
\begin{equation}\label{def_l}
\L(z)^2\to \l(z)^2=\prod_{l=1}^{\Nc}\dfrac{z-\xi_l+\e_1}{z-\xi_l}\prod_{r=1}^{\NB}\dfrac{(z-u_r)^2-\e_1^2}{(z-u_r)^2}.
\end{equation}
The first factor tends to one if we send the cut-offs $n_l\to\infty$.

\subsection{NS limit of the algebra}
In the NS limit, we rescale the generators $D_{m,n}$ to define $d_{\pm n,m}=\e_2^{n/2+m}D_{\pm n,m}$ with $n,m\geq 0$. It implies that the limit of the generating series \ref{def_D} reads
\begin{equation}\label{d0_series}
\e_2^{1/2}D_{\pm1}(z)\to d_{\pm1}(z)=\sum_{n\geq0}(z+\e_1)^{-n-1}d_{\pm1,n},\quad \e_2D_0(z)\to d_0(z)=\sum_{n\geq1}(z+\e_1)^{-n}d_{0,n}.
\end{equation}
The commutation relations \ref{comm_SHc} become
\begin{equation}\label{comm_d}
[d_0(z),d_{\pm1}(w)]=\pm\e_2\dfrac{d_{\pm1}(w)-d_{\pm1}(z)}{z-w},\quad [d_{-1}(z),d_1(w)]=\e_2\dfrac{e(w)-e(z)}{z-w},
\end{equation}
with $e_n=\e_2^nE_n$ and $E(z)\to e(z)$. To obtain the relation between modes $e_n$ and $d_{0,n}$ from \ref{rel_E_D0} we need to take the limit of the functions $\pi_n(s)$ and $\o_n(s)$ with $\e_2/s$ fixed.\footnote{We find for $\eta=\e_2/s=z+\e_1$:
\begin{align}
\begin{split}
&\pi_n(s)\to\e_2^n\eta^{-n}G_n(1+\e_1/\eta),\\
&\o_n(s)\to -(n+1)\e_2^{n+1}\eta^{-n-1}\left[G_{n+1}(1+\e_1/\eta)+G_{n+1}(1-\e_1/\eta)\right].
\end{split}
\end{align}}
We deduce
\begin{equation}\label{rel_e_d0_II}
1+\e_1e(z)=\Pi(z)\exp\left(-\sum_{n\geq1}n(z+\e_1)^{-n}\left[G_n(1+\e_1/(z+\e_1))+G_n(1-\e_1/(z+\e_1))\right]d_{0,n}\right),
\end{equation}
with the function $\Pi(z)$ depending on the finite rescaled central charges $\e_2^nc_n$:
\begin{equation}\label{def_Pi}
\Pi(z)=\exp\left(\sum_{n\geq0}(-1)^{n+1}\e_2^nc_n(z+\e_1)^{-n}G_n(1+\e_1/(z+\e_1))\right)=\prod_{l=1}^{\Nc}\dfrac{z+\e_1-a_l}{z-a_l}.
\end{equation}

The current and Virasoro modes scale as
\begin{equation}
J_{\pm n}=(-\sqrt{-\e_1})^{-n}d_{\mp n,0},\quad \e_2 L_{\pm n}=\dfrac1n(-\sqrt{-\e_1})^{-n}d_{\mp n,1}+\hf(1-n)N\e_1J_{\pm n}.
\end{equation}
The rescaling of the Liouville field by $2b$ in the semiclassical limit indeed brings a factor $\e_2$ to the Virasoro modes since $T_\text{cl}(z)=4b^2T_\text{qu}(z)$ and $b\sim \sqrt{\e_2}$. We note that most of the correlators are now vanishing at leading order in $\e_2$, including $[d_{\pm1}(z),d_{\pm 1}(w)]=O(\e_2)$. This is to be expected since $\e_2$ plays the role of $\hbar\sim b^2$ in the semiclassical limit of Liouville theory. In this limit, a Virasoro algebra with $c=1$ is still present, but the commutators must be replaced by Poisson brackets \cite{Teschner2001}. It would be extremely interesting to investigate this phenomenon more deeply, but it is out of the scope of this paper.

Our choice of rescaling for $d_{\pm1,n}$ renders the KMZ transformation \ref{KMZ} finite,
\begin{equation}\label{KMZ_NS}
\sqrt{-\e_1}\d_{\pm1,z}\CZ[Y_1,Y_2]=u_{\pm1,z}\CZ[Y_1,Y_2],
\end{equation}
where we absorbed the factor $\sqrt{\e_2}$ in the definition of $\d_{\pm1,z}$, and the functions defined in \ref{def_U} simplifies into $U_{\pm1,z}\to u_{\pm1,z}$ with:
\begin{align}
\begin{split}\label{def_upm}
&u_{-1,z}=\prod_{l=1}^{\Nc}\dfrac{z+m+\e_1/2-\x_l^{(2)}}{z-\x_l^{(1)}}\prod_r\dfrac{z-\e_1-u_r}{z-u_r}\prod_s\dfrac{z+m+\e_1/2-v_s}{z+m-\e_1/2-v_s}-1,\\
&u_{1,z}=\prod_{l=1}^{\Nc}\dfrac{z+\e_1-\x_l^{(1)}}{z+m+\e_1/2-\x_l^{(2)}}\prod_r\dfrac{z+\e_1-u_r}{z-u_r}\prod_s\dfrac{z+m-\e_1/2-v_s}{z+m+\e_1/2-v_s}-1.
\end{split}
\end{align}
The factors containing $\x_l^{(a)}$ cancel each-other in the limit $n_l^{(a)}\to\infty$.

The SHc algebra is rather disappointing in the NS limit since all commutators vanish. It is however possible to obtain non-trivial commutation relations under the rescaling of $d_0(z)$ by $\e_2^{-1}$ and $d_{\pm1}(z)$ by $\e_2^{-1/2}$:
\begin{equation}
[d_0(z),d_{\pm1}(w)]=\pm\dfrac{d_{\pm1}(w)-d_{\pm1}(z)}{z-w},\quad [d_{-1}(z),d_1(w)]=\dfrac{e(w)-e(z)}{z-w}.
\end{equation}
The commutation relations for $[d_{\pm1}(z),d_{\pm1}(w)]$ will also be of order one. Note however that the relation \ref{rel_e_d0_II} between $e(z)$ and $d_{0,n}$ generators simplifies: the RHS becomes $\Pi(z)$ which is independent of the $d_{0,n}$ generators: $e_n$ are now simple central charges. We expect this algebra to be realized by some operators of the integrable system. In that respect, it is satisfying to see that the infinite number of modes $d_{0,n}$ remain commuting, and might be identified with the conserved charges built upon Dunkl operators in qNLS.

\subsection{Covariance of the bifundamental contribution under SHc transformations}
The previous description of the NS limit for the objects $Y_a$ allows to derive the limit of the bifundamental contribution $\CZ[Y_1,Y_2]$ defined in \ref{CZ_YY}. This quantity depends on two sets of Bethe roots, and will be denoted $z[u,v]$. The derivation is a bit lengthy, it is done in the appendix \refOld{App_BZ}. The main technicality is to get rid of the dependence in the dual variables $\tilde{t}_{l,j}$. The final expression is\footnote{It still contains a trivial $\e_2$ dependence since $\e_2\log\CZ[Y_1,Y_2]=O(1)$ as $\e_2\to0$.}
\begin{align}
\begin{split}\label{zuv}
z[u,v]&=\prod_{r=1}^{N_B^{(1)}}\prod_{l=1}^{\Nc}\dfrac{g(u_r-\xi_l^{(2)}+m+\e_1/2)}{\sqrt{g(u_r-\xi_l^{(1)})g(u_r-\xi_l^{(1)}+\e_1)}}\prod_{r=1}^{N_B^{(2)}}\prod_{l=1}^{\Nc}\dfrac{g(v_r-\xi_l^{(1)}-m+\e_1/2)}{\sqrt{g(v_r-\xi_l^{(2)})g(v_r-\xi_l^{(2)}+\e_1)}}\\
&\times \prod_{r,s=1}^{N_B^{(1)}}\left(\dfrac{g(u_r-u_s-\e_1)}{g(u_r-u_s+\e_1)}\right)^{1/4}\times\prod_{r,s=1}^{N_B^{(2)}}\left(\dfrac{g(v_r-v_s-\e_1)}{g(v_r-v_s+\e_1)}\right)^{1/4}\times\prod_{r=1}^{N_B^{(1)}}\prod_{s=1}^{N_B^{(2)}}\dfrac{g(u_r-v_s+m+\e_1/2)}{g(u_r-v_s+m-\e_1/2)}.
\end{split}
\end{align}
This expression involves the function $g(x)$ defined as $g(x)=x^{x/\e_2}$. This function has a branch cut on the negative real axis, such that $g(e^{2i\pi} x)=e^{2i\pi x/\e_2} g(x)$. It implies that the expression we provided for $z[u,v]$ is ambiguous, and the sheet where $x$ lies in $g(x)$ must be specified. In the next section, the logarithm of $z[u,v]$ will be related to a Yang-Yang functional that involves, in addition to the two sets of Bethe roots $u_r^{(a)}$ ($u_r^{(1)}=u_r$, $u_r^{(2)}=v_r$), two sets of integers $\eta_r^{(a)}$ that can be associated to the sheets of the function $g(x)$. We will further justify that this ambiguity is irrelevant for the KMZ transformation.

The function $g(x)$ satisfies the property $g(x\pm\e_2)\simeq g(x)x^{\pm1} e^{\pm1}$ at first order in $\e_2$. It allows to study the variation of $z[u_r,v_s]$ under an infinitesimal shift of one Bethe root: $u_r\to u_r\pm\e_2\d_{r,r'}$. The constant factors $e^{\pm1}$ cancel each other in the ratios, and we get
\begin{equation}\label{shift_z}
\dfrac{z[u_r\pm\e_2\d_{r,r'},v_s]}{z[u_r,v_s]}=\prod_{l=1}^{\Nc}\dfrac{(u_{r'}-\xi_l^{(2)}+m+\e_1/2)^{\pm1}}{\left[(u_{r'}-\xi_l^{(1)})(u_{r'}-\xi_l^{(1)}+\e_1)\right]^{\pm\frac12}}\prod_{\superp{s'=1}{s'\neq {r'}}}^{N_B^{(1)}}\left(\dfrac{u_{r'}-u_{s'}-\e_1}{u_{r'}-u_{s'}+\e_1}\right)^{\pm\frac12}\prod_{s'=1}^{N_B^{(2)}}\left(\dfrac{u_{r'}-v_{s'}+m+\e_1/2}{u_{r'}-v_{s'}+m-\e_1/2}\right)^{\pm1}.
\end{equation}
A similar expression is obtained by considering shifts of the Bethe roots $v_s$, exchanging the role of $u_r$ and $v_s$ and flipping the sign of the bifundamental mass.

From now on, we send the cut-offs $n_l^{(a)}=L\to\infty$ such that the first product in the previous expression vanishes. In order to define the SHc generators, we need to introduce a decomposition of the function $\l(z)^2$ as a sum over its poles, as we did for $\L(z)^2$ in \ref{prop_L}:
\begin{equation}\label{def_lr}
\l(z)^2=1+\e_1\sum_r\dfrac{\p}{\p u_r}\left(\dfrac{\l_r^2}{z-u_r}\right),\quad \l_r^2=-\e_1\prod_{s\neq r}\dfrac{(u_r-u_s)^2-\e_1^2}{(u_r-u_s)^2}.
\end{equation}
The factors $\l_r$ play a role equivalent to $\L_x(Y)$ in the NS limit since $\L_x(Y)\simeq \l_r/\sqrt{\e_2}$. The information of the states $|Y>$ is encoded in the Young diagrams structure. In the NS limit, this information is rendered by the knowledge of the set of Bethe roots $\{u_r\}$. It is then natural to introduce a state $|u>$ which is an eigenstate of $d_0(z)$ with the eigenvalue
\begin{equation}\label{d0_ur}
d_0(z)|u>=\left(-\sum_{r=1}^{N_B}\log(z-u_r)+\sum_{l=1}^{\Nc}\sum_{i=1}^{n_l}\log(z-a_l-(i-1)\e_1)\right)|u>.
\end{equation}
This definition of the action of $d_0(z)$ will become clear in the next section. The second term in the RHS acts as a regulator to ensure $d_0(z)\simeq \e_2|Y|/z$ at infinity.

The generators constituting $e(z)$ are also diagonal in the basis $|u>$, with
\begin{equation}
\left[1+\e_1e(z)\right]|u>=\l(z)^2|u>
\end{equation}
inherited from \ref{vp_E0}. Finally, the action of $D_{\pm1}(z)$ involves the addition or subtraction of boxes to the Young diagrams. These variations of the columns height generates shifts of the Bethe roots. Using an exponential notation for shift operators, we define
\begin{equation}\label{d1_ur}
d_{\pm1}(z)=\sum_re^{\pm\e_2\p_r}\dfrac{\hat\l_r}{z-\hat u_r},\quad \p_r=\dfrac{\p}{\p u_r}.
\end{equation}
In this definition, we employed the 'Dunkl' operator $\hat u_r$ such that $\hat u_r|u>=u_r|u>$. It satisfies with the shift operators the commutation relation $[\hat u_r,e^{\pm\e_2\p_r}]=\pm\e_2e^{\pm\e_2\p_r}$. The operator $\hat\l_r$ is obtained from $\l_r$ by replacing $u_r$ with $\hat u_r$, it satisfies $[\hat\l_r,e^{\pm\e_2\p_s}]=\pm\e_2e^{\pm\e_2\p_s}\p_s\hat\l_r$. An explicit calculation shows that the operators $d_{\pm1}(z),d_0(z)$ obey the commutation relations \ref{comm_d} of SHc at first order in $\e_2$. In addition, expanding over $z$, one can show that the generators $d_{0,n}$ and $e_n$ are related as in \ref{rel_e_d0_II}.\footnote{The expansion of $d_0(z)$ gives
\begin{equation}
d_{0,n}|u>=\dfrac1n\sum_{r=1}^{N_B}[(u_r+\e_1)^n-(\bx_r+\e_1)^n]|u>,
\end{equation}
with $\bx_r=a_l+(i-1)\e_1$ for $r=(l,i)$ with $l=1\cdots \Nc$ and $i=1\cdots n_l$. This expression can be plugged into \ref{rel_e_d0_II}, and the summation over $n$ performed with the help of \ref{form_Gn}. It reproduces the function $\l(z)^2$.}

The action of $d_{\pm1}(z)$ on the bifundamental contributions is obtained from \ref{shift_z}. In this expression, one recognize the decomposition of $u_{\pm1,z}$ over the poles at $z=u_r$. Defining the KMZ transformation as in \ref{KMZ_CZ}:
\begin{equation}\label{def_pm1}
\d_{\pm1,z}=\mp d_{\mp1}(z)|_{u_r}\pm d_{\pm1}(z+m\pm\e_1/2)|_{v_r},
\end{equation}
where we indicated on which set of Bethe roots the operator is actually acting, we recover exactly the KMZ transformation \ref{KMZ_NS}. It is interesting to note that in order to satisfy the SHc commutation relations, it is sufficient to expand in $\e_2$ the shift operators in the definition of $d_{\pm1}(z)$, and keep only the terms of order $O(\e_2)$. However, the full exponential is needed to obtain the KMZ transformation.

\section{Connections with the Mayer cluster expansion}
\subsection{Integral expression of the $A_2$ partition function}
The instanton partition function of $\mathcal{N}=2$ gauge theories has been originally obtained as a set of coupled contour integrals \cite{Nekrasov2003}. In the case of the $A_2$ quiver,
\begin{equation}\label{def_CZA2}
\CZ_{A_2}=\sum_{N_1,N_2=0}^\infty\dfrac{\tilde{q}_1^{N_1}\tilde{q}_2^{N_2}}{N_1!N_2!}\ \int{\prod_{i=1}^{N_1}\prod_{j=1}^{N_2}K_{12}(\phi_{i,1}-\phi_{j,2})\prod_{a=1,2}\left(\prod_{\superp{i,j=1}{i<j}}^{N_a}K(\phi_{i,a}-\phi_{j,a})\prod_{i=1}^{N_a}Q_a(\phi_{i,a})\dfrac{d\phi_{i,a}}{2i\pi}\right)},
\end{equation}
with $\tilde{q}_a=\e_+q_a/\e_1\e_2$. The integration contours over the instanton positions $\phi_{i,a}$ lie along the real axis. They are closed in the upper half plane but avoid possible singularities at infinity. The potential attached to a node $a$ is a ratio involving the mass polynomial $m_a(x)$ and the gauge polynomial $A_a(x)$. These monic polynomials have zeros respectively at the value of the masses $m_f^{(a)}$ of fundamental matter fields, and at the Coulomb branch vevs $a_l^{(a)}$:
\begin{equation}\label{pot_quiver}
Q_a(x)=\dfrac{m_a(x)\prod_{b\neq a}A_b(x+\e_+/2+m_{ab})}{A_a(x)A_a(x+\e_+)},\qquad\text{with}\quad m_a(x)=\prod_{f=1}^{\Nc}(x-m_f^{(a)}),\quad A_a(x)=\prod_{l=1}^{\Nc}(x-a_l^{(a)}).
\end{equation}
As in the first section, we have assumed that both nodes have a gauge group $U(N_c)$ and $N_c$ fundamental flavors. The kernel $K$ is associated to the nodes, and $K_{12}$ to the arrow $1\to2$. The latter depends on the bifundamental mass $m=m_{12}=-m_{21}$. The expressions of the kernels $K$ and $K_{12}$ can be found in \cite{Shadchin2005,Meneghelli2013}, they are reproduced in the appendix \refOld{AppA}. They both depend on the $\Omega$-background equivariant parameters $\e_1,\e_2$ which are assumed purely imaginary. Coulomb branch vevs have a positive imaginary part, and fundamental masses $m_f^{(a)}$ are real.

The integrals can be evaluated as a sum over residues, and poles are known to be in one to one correspondence with the boxes of the set of objects $Y_a$ through the map \ref{map_phix}. The resulting expression for the $A_2$ quiver has been given in \ref{decomp_CZ}. However, the integral definition reveals more convenient for taking the NS limit. This procedure has been described in \cite{Meneghelli2013,Bourgine2014}. We employ here the method of \cite{Bourgine2014}, which has been performed only in the case of a single gauge group. It is easily extended to arbitrary quivers, this is done in the appendix \refOld{AppA}.

In the appendix \refOld{AppA}, two different expressions of the partition function in the NS limit are provided (equ \ref{SGC_onshell} and \ref{cluster_hadron}). The first one will be studied in the next subsection, it reproduces the familiar form of the Yang-Yang functional \cite{Yang1968}. The underlying Bethe equations are related to those derived in the third section, thus bringing a justification for the assumption made on the shape of Young diagrams in the NS limit. The bifundamental contribution $z[u,v]$ given in \ref{zuv} can be recovered from the Yang-Yang functional, and we will be able to comment on its ambiguities. The second expression for the NS partition function is derived in the appendix \refOld{AppA2}, and will be studied in the subsection \refOld{sec_43}. It consists of a sum of coupled integrals reminding of a matrix model. The integration variables are interpreted as the position of hadrons in a one dimensional space. They can be formally related to the Bethe roots, and we will deduce the eigenvalues \ref{d0_ur} of the operator $d_0(z)$.

\subsection{NLIE, Bethe roots and Yang-Yang potential}
In \cite{Nekrasov2009}, it was claimed that the free energy at first order in $\e_2$ can be expressed as an on-shell effective action. This result is derived for the $A_2$ quiver in the appendix \refOld{AppA1}. The action \ref{SGC_onshell} depends on two fields indexed by the node $a$: the density $\rho_a(x)$ and the pseudo-energy $\ve_a(x)$. The equations of motion relate these two fields as
\begin{equation}\label{def_ve}
2i\pi\rho_a(x)=-\log\left(1-q_ae^{-\ve_a(x)}\right),
\end{equation}
and produce for the fields $\ve_a(x)$ a set of NLIE,
\begin{equation}\label{NLIE_quiver}
\ve_a(x)+\log Q_a(x)-\sum_b\int{G_{ab}(x-y)\log(1-q_be^{-\ve_b(y)})\dfrac{dy}{2i\pi}}=0,
\end{equation}
with the kernels $G_{11}(x)=G_{22}(x)=G(x)$ and $G_{21}(x)=G_{12}(-x)$,
\begin{equation}\label{G11}
G(x)=\p_x\log\left(\dfrac{x+\e_1}{x-\e_1}\right),\quad G_{12}(x)=\p_x\log\left(\dfrac{x+m-\e_1/2}{x+m+\e_1/2}\right).
\end{equation}
In the NLIE \ref{NLIE_quiver}, the integration is done over the same contour as in the original definition \ref{CZ_A2} of the partition function $\CZ_{A_2}$.

The set of NLIE \ref{NLIE_quiver} is characteristic of the Thermodynamical Bethe Ansatz technique developed in \cite{Yang1968,Zamolodchikov1990a}. It is usually associated to the following system of Bethe equations,
\begin{align}
\begin{split}\label{Bethe_2}
&q_1Q_1(\bu_r)\prod_{s=1}^{\NBb^{(1)}}\dfrac{\bu_r-\bu_s-\e_1}{\bu_r-\bu_s+\e_1}\prod_{s=1}^{\NBb^{(2)}}\dfrac{\bu_r-\bv_s+m+\e_1/2}{\bu_r-\bv_s+m-\e_1/2}=1,\\
&q_2Q_2(\bv_r)\prod_{s=1}^{\NBb^{(2)}}\dfrac{\bv_r-\bv_s-\e_1}{\bv_r-\bv_s+\e_1}\prod_{s=1}^{\NBb^{(1)}}\dfrac{\bv_r-\bu_s-m+\e_1/2}{\bv_r-\bu_s-m-\e_1/2}=1.
\end{split}
\end{align}
Indeed, the method elaborated in \cite{Destri1994} shows that the counting functions $\eta_a(x)$ obey the NLIE \ref{NLIE_quiver} with an appropriate contour of integration.\footnote{Counting functions $\eta_a(x)$ are defined such that they take integer values at the Bethe roots $x=\bu_r^{(a)}$, for instance
\begin{equation}
2i\pi\eta_1(x)=\log\left(q_1Q_1(x)\prod_{s=1}^{\NBb^{(1)}}\dfrac{x-\bu_s-\e_1}{x-\bu_s+\e_1}\prod_{s=1}^{\NBb^{(2)}}\dfrac{x-\bv_s+m+\e_1/2}{x-\bv_s+m-\e_1/2}\right).
\end{equation}
They are identified with the pseudo-energy as $2i\pi\eta_a(x)=-\ve_a(x)+\log q_a$.} This method is briefly reviewed in \cite{Bourgine2014}. In particular, it was observed that the derivative of the field $\rho_a(x)$ correspond to a density of Bethe roots:\footnote{Since we are dealing with contour integrals, the definition of the delta function may require some clarification. Here, it is defined as the operator
\begin{equation}
\int\d(x-y)f(y)dy=f(x)
\end{equation}
for $x$ inside the integration contour (i.e. $\Im x>0$) which will always be the case.}
\begin{equation}\label{id_densities}
-\dfrac{d}{dx}\rho_a(x)=\brho_{B}^{(a)}(x), \quad\text{for}\quad \brho_B^{(a)}(x)=\sum_{r=1}^{\NB^{(a)}}\d(x-\bu_r^{(a)}),
\end{equation}
with the shortcut notations $\bu_r^{(1)}=\bu_r$ and $\bu_s^{(2)}=\bv_s$.

Comparing the Bethe equations \ref{Bethe_2} with those obtained in the second section, we observe that they match only in the formal case $\Nc=0$. The mismatch comes from missing factors in the node potential \ref{Bethe_1}. However, if we decompose the set of Bethe roots $\{u_r^{(a)}\}$ into a set of fixed variables $\{\bar{x}_{l,i}^{(a)}=a_l^{(a)}+(i-1)\e_1\}$ with $i=1\cdots n_l^{(a)}$ and $l=1\cdots\Nc$, and a set of free variables $\{\bu_r^{(a)}\}$, then we observe that the Bethe equations \ref{Bethe_1} for $\{u_r^{(a)}\}$ become the Bethe equations \ref{Bethe_2} for $\{\bu_r^{(a)}\}$. Thus, to each solution $\{\bu_r,\bv_s\}$ of the Bethe equations \ref{Bethe_2} corresponds a solution $\{u_r,v_s\}$ of \ref{Bethe_1} with the additional roots given by $\{\bar{x}_{l,i}^{(1)},\bar{x}_{l',i'}^{(2)}\}$.

It results from this identification that the relation \ref{id_densities} among densities is modified for the Bethe roots $u_r^{(a)}$,
\begin{equation}\label{id_densities_II}
-\dfrac{d}{dx}\rho_a(x)=\rho_B^{(a)}(x)-\rho\pert_a(x),\quad\text{for}\quad\rho_B^{(a)}(x)=\sum_{r=1}^{\NB^{(a)}}\d(x-u_r^{(a)}),\quad \rho\pert_a(x)=\sum_{l=1}^{\Nc}\sum_{i=1}^{n_l^{(a)}}\d(x-\bx_{l,i}^{(a)}).
\end{equation}
Inserted in the integrals of \ref{NLIE_quiver} through the relation \ref{def_ve}, this amended identity gives an expression of the pseudo-energy in terms of the Bethe roots $u_r^{(a)}$ where $Q_a(x)$ is replaced with $m_a(x)\xi_a(x)$.\footnote{We use the property
\begin{equation}
\left(1-e^{-\e_1\p_x}\right)\rho\pert_a(x)=\sum_{l=1}^{\Nc}\left[\d(x-a_l^{(a)})-\d(x-\xi_l^{(a)})\right].
\end{equation}} It allows to identify $2i\pi\eta_a(x)=-\ve_a(x)+\log q_a$ with the counting function associated to the Bethe equations \ref{Bethe_1}. Those equations are obtained by imposing $\eta_a(u_r^{(a)})\in\mathbb{Z}$, i.e. $e^{2i\pi \eta_a(u_r^{(a)})}=1$. Interestingly, the correction term in \ref{id_densities_II} reproduces the part of the density involved in the perturbative contributions to the gauge theory partition function \cite{Bourgine2012a}. It allows to interpret the Bethe equations \ref{Bethe_1} of the second section as the extremum of the full $\mathcal{N}=2$ partition function, including the perturbative contribution.\footnote{We refer here to the case of Young diagrams with infinitely many columns, such that the factor $\xi_a(x)$ is not present.} On the other hand, the Bethe equations \ref{Bethe_2} obtained here describe the extremization of the instanton contribution only.

We now examine the free energy $\CF_{A_2}=\lim_{\e_2\to0} \e_2\log \CZ_{A_2}$ derived from the on-shell action \ref{SGC_onshell}. Instead of the pseudo-energies $\ve_a(x)$, we prefer to work with the counting functions $\eta_a(x)$,
\begin{equation}
\CF_{A_2}=\sum_{a=1,2}\CF_a[\rho_a]+\CF_{12}[\rho_1,\rho_2]+\sum_{a=1,2}\SJ[\rho_a,\eta_a].
\end{equation}
In the RHS, we have singled out the parts that depend on $\eta_a$:
\begin{equation}\label{CS_rho_eta}
\SJ[\rho,\eta]=-2i\pi\int{\rho(x)\eta(x)dx}+\dfrac1{2i\pi}\int{\text{Li}_2\left(e^{2i\pi\eta(x)}\right)dx},
\end{equation}
and decomposed the remaining term into three contributions:
\begin{align}
\begin{split}
\CF_a[\rho_a]&=\dfrac14\int{\rho_a(x)\rho_a(y)G(x-y)dxdy}+\int{\rho_a(x)\log\left(\dfrac{q_am_a(x)}{\sqrt{A_a(x)A_a(x+\e_1)}}\right)}dx,\\
\CF_{12}[\rho_1,\rho_2]&=\dfrac14\sum_{a=1,2}\int{\rho_a(x)\rho_a(y)G(x-y)dxdy}+\int{\rho_1(x)\rho_2(y)G_{12}(x-y)dxdy}\\
&+\sum_{a=1,2}\int{\rho_a(x)\log\left(\dfrac{A_{b\neq a}(x+m_{ab}+\e_1/2)}{\sqrt{A_a(x)A_a(x+\e_1)}}\right)}dx.
\end{split}
\end{align}
In these expressions, the densities $\rho_a(x)$ can be replaced by the Bethe roots density $\rho_B^{(a)}(x)$ using an integration by parts and the relation \ref{id_densities_II}, leading to summations over Bethe roots $u_r$ and $v_s$. In particular, the exponential of $\CF_{12}[\rho_1,\rho_2]$ reproduces the expression found in \ref{zuv} for $z[u,v]$,
\begin{equation}
e^{\frac1{\e_2}\CF_{12}[\rho_1,\rho_2]}\equiv z[u,v],
\end{equation}
up to ambiguities of the same nature as in section three. Similarly, $\CF_a[\rho_a]$ reproduces the NS limit of the matter bifundamental contributions $\CZ[M_a,Y_a]$. It is noted that, if there are several solutions to the Bethe equations, the partition function is a sum over the contributions of each solution.

We have thus shown that $\CF_{12}$ and $\CF_a$ reproduce the partition function of the $A_2$ quiver derived in the second section. But we have here an additional term $\SJ[\rho,\eta]$ that remains to be treated. To do so, we repeat the trick used in \cite{Bourgine2014}: in \ref{CS_rho_eta}, the second integral simplifies using an integration by parts and the relation between the density and the counting function derived from \ref{def_ve}. One of the terms obtained cancel with the first term in the expression of $\SJ[\rho,\eta]$, and it only remains
\begin{equation}
\SJ[\rho,\eta]=2i\pi\int{x\rho'(x)\eta(x)dx}.
\end{equation}
For the first node, it gives from \ref{id_densities_II} in terms of the Bethe roots:
\begin{equation}
\SJ[\rho_1,\eta_1]=-2i\pi\sum_ru_r\eta_r^{(1)}+2i\pi\sum_{l=1}^{\Nc}\sum_{i=1}^{n_l^{(1)}}\bx_{l,i}^{(1)}\eta_1(\bx_{l,i}^{(1)}).
\end{equation}
with $\eta_r^{(1)}=\eta_1(u_r)\in\mathbb{Z}$. The second term in the RHS is independent of the Bethe roots, and simply produces an overall constant factor. The first term is more interesting since it has the same form as the ambiguity arising in \ref{zuv} from $2i\pi$ rotations of the argument of the functions $g(x)$. Since $\eta_r^{(1)}$ is an integer, a shift of $u_r\to u_r\pm\e_2$ leads to a variation $\d\SJ=\mp2i\pi\e_2\eta_r^{(1)}$ which vanishes when exponentiated with a factor $1/\e_2$. Thus, this ambiguity does not interfere with the KMZ transformation we have defined.

The full expression $\CF_{A_2}[u_r^{(a)},\eta_r^{(a)}]$ reproduces the Yang-Yang functional of qNLS, up to the potential term. Thus, \ref{d0_ur} and \ref{d1_ur} define an action of the SHc operators at first order in $\e_2$ on the Yang-Yang functional of an integrable system.

\subsection{Hadronic integrals}\label{sec_43}
Another expression of the partition function $\CZ_{A_2}$ in the NS limit has been obtained in the appendix \refOld{AppA2},
\begin{equation}\label{CZ_A2}
\CZ_{A_2}\simeq\sum_{p,q=1}^{\infty}\dfrac{\e_2^{-p-q}}{p!q!}\sum_{k_1,\cdots k_p=1}^{\infty}\sum_{l_1,\cdots,l_q=1}^\infty \dfrac{q_1^{\sum_ik_i}q_2^{\sum_jl_j}}{\prod_ik_i^2\prod_jl_j^2}\int{\prod_{i=1}^p\dfrac{dx_i}{2i\pi}\prod_{j=1}^q\dfrac{dy_j}{2i\pi}\mathscr{I}(k,x|l,y)},
\end{equation}
with the integrand
\begin{align}
\begin{split}\label{def_z_kx_ly}
\mathscr{I}(k,x|l,y)&=\prod_iQ_1(x_i)^{k_i}\prod_jQ_2(y_j)^{l_j}\exp\left(\e_2\sum_{i,j}k_il_jG_{12}(x_i-y_j)\right)\\
&\times \exp\left(\frac{\e_2}2\sum_{i,j}k_ik_jG(x_i-x_j)+\frac{\e_2}2\sum_{i,j}l_il_jG(y_i-y_j)\right).
\end{split}
\end{align}
This expression is a consequence of a phenomenon of \textit{clustering} or \textit{confinement} of the instantons as $\e_2\to0$. Indeed, the integrations in the original expressions \ref{def_CZA2} are done over the positions of $N_1+N_2$ instantons denoted $\phi_{i,a}$. It is common in the study of random matrix models to see the integration variables $\phi_{i,a}$ as pseudo-particles of color $a$, in an external potential $\log Q_a$, and interacting through the kernels $K$ and $K_{12}$. In the NS limit, a number $k_i$ of instantons at positions $\phi_{\a,1}$ ($\a=1\cdots k_i$) may become very close to each other due to the infinitely strong short-range interaction of $K(x)$. These instantons, at a distance $\sim\e_2$ of each-other, can be approximated by a single particle of \textit{charge} $k_i$ and \textit{center of mass} $x_i$. By analogy with quarks bound states, it will be called a \textit{hadron}. This phenomenon is described with more details in \cite{Bourgine2014}. The factors $1/k_i^2$ in the summation \ref{CZ_A2} are a consequence of the interaction between the elementary instantons composing a hadron. For the second node, hadrons positions are denoted $y_j$ and they have the charge $l_j$.

The equality \ref{CZ_A2} holds at leading order in $\e_2$. Both sides contain $\e_2$ corrections that do not match. To extract the leading order in $\e_2$, (generalized) matrix model techniques can be employed \cite{Bourgine2013}. A Mayer expansion can also be considered, the first order free energy being expressed as a sum over connected clusters with a tree structure, and vertices associated to hadrons. The free energy in the NS limit is equal to the Legendre transform of the canonical free energy at large $N_a$, where $N_a$ is the conjugate variable of the chemical potential $\log q_a$. In this setting, only configurations $(k,x)$ with $\e_2\sum_ik_i=O(1)$ contribute to the leading order. The charges $k_i$ are supposed to be finite, and the sum is over $i\in\mathbb{Z}^{>0}$. Thus, we will consider that for any test function $f(x)$, the following summations are of order one:
\begin{equation}\label{sum_kfx}
\e_2\sum_{i=1}^\infty k_if(x_i).
\end{equation}

\subsubsection{Hadronic variables and Bethe roots}
In the hadronic formulation, the pseudo-energy is seen as a Lagrange multiplier enforcing the following identities,
\begin{equation}\label{cond_density}
\rho_1(x)=\e_2\sum_{i=1}^p{k_i\d(x-x_i)},\quad \rho_2(y)=\e_2\sum_{j=1}^q{l_j\d(y-y_j)},
\end{equation}
where the field $\rho_a(x)$ is defined as the dressed vertex of the Mayer expansion. Although pertaining to different contexts, it is instructive to formally identify these densities with the fields of the previous subsection, and use \ref{id_densities_II} to provide a bridge between the expressions in terms of $(k_i,x_i)$ and $u_r$ variables.

It is possible to give another argument in favor of the relation \ref{id_densities_II} between variables $(k_i,x_i)$ and $u_r$. Let $f(x)$ be a test function without singularities in the upper half plane.\footnote{Typically $f(x)$ is a polynomial in $x$, or the generating function $1/(z-x)$ for $\Im z<0$.} The summation \ref{sum_kfx} of $k_if(x_i)$ can be interpreted as an operator evaluated in the statistical average that defines the NS partition function \ref{CZ_A2}. As such, it is the NS limit of the operator
\begin{equation}
\dfrac{\e_1\e_2}{\e_+}\sum_if(\phi_{i,1})
\end{equation}
evaluated as the vev associated to the original partition function \ref{def_CZA2}. This vev can also be computed as a sum over residues, and since $f(x)$ has no singularities within the contour of integration, we find
\begin{equation}
\dfrac{\e_1\e_2}{\e_+}\sum_{x\in Y_1}f(\phi_x)
\end{equation}
with the object $Y_1$ defined in the first section, and the map $x\to\phi_x$ in \ref{map_phix}. Thus, in the NS limit, we can formally replace
\begin{equation}\label{dico_kx}
\dfrac{\e_1\e_2}{\e_+}\sum_{x\in Y_1}f(\phi_x)\to\e_2\sum_ik_if(x_i),
\end{equation}
and similarly for $Y_2$ and the variables $(l_j,y_j)$. The consistency with the identification of Bethe roots as the position of boxes on Young diagrams edges \ref{dico_ur}\footnote{As an example, we can use the identity \ref{rel_r_L} that provides an alternative expression of the function $\L(z)^2$ as a product over the boxes in $Y$. As $\e_2\to0$, the function $r(z)$ defined in \ref{def_rz} is expended as
\begin{equation}\label{r_NS}
r(z)=1+\e_2\th(z)+O(\e_2^2),\quad \theta(z)=\p_z\log\left(\dfrac{z^2}{z^2-\e_1^2}\right).
\end{equation}
Performing the NS limit of the LHS of \ref{rel_r_L} as in \ref{dico_kx}, and expressing the RHS with \ref{def_l} in terms of Bethe roots, we get
\begin{equation}\label{id_for_l(z)}
\prod_{r=1}^{N_B^{(1)}}\dfrac{(z-u_r)^2-\e_1^2}{(z-u_r)^2}=\prod_{l=1}^{\Nc}\dfrac{(z+\e_1-a_l)(z-\xi_l)}{(z-a_l)(z+\e_1-\xi_l)}e^{\e_2\sum_i{k_i\th(z-x_i)}},
\end{equation}
in agreement with \ref{id_ur_kx}. Note that this identity has a well defined limit when $\xi_l\to\infty$.}
\begin{equation}\label{id_ur_kx}
\sum_{r=1}^{N_B}f(u_r)=\e_2\sum_ik_if'(x_i)+\sum_{l=1}^{\Nc}\sum_{i=1}^{n_l}f(\bx_{l,i}).
\end{equation}
This is the equivalent of the relation \ref{id_densities_II} for the densities. 

\begin{figure}[!t]
\centering
\includegraphics[width=9cm]{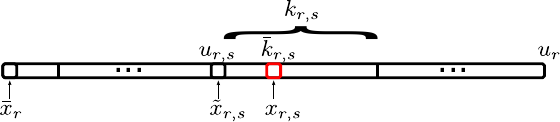}
\caption{Decomposition of an infinite column (here rotated by 90\textdegree) into finite sets of $k_{r,s}$ boxes.}
\label{fig_d3}
\end{figure}

There is a simple heuristic interpretation of the relation we have observed between the variables $(k_i,x_i)$ and $u_r$. It boils down to represent the Young tableaux columns as made of an infinite number of hadrons. For simplicity, let us assume $\Nc=1$ with a unique Coulomb branch vev $a_1=a$, and $N_B=n_1=n$. The generalization to $N_c>1$ is straightforward. We drop the color and node indices, and consider the expression
\begin{equation}
\sum_{r=1}^{N_B}[f(u_r)-f(\bx_r)]=\sum_{r=1}^n [f(\bx_r+\l_r\e_2)-f(\bx_r)],
\end{equation}
due to the identification of $u_r$ with $t_{1,r}$, and $\bx_r=\bx_{1,r}=a+(r-1)\e_1$. The difference of the two functions in the RHS can be replaced by an integral of the derivative,
\begin{equation}
\sum_{r=1}^{N_B}[f(u_r)-f(\bx_r)]=\sum_{r=1}^n\int_0^{\l_r}{f'(\bx_r+\e_2 t_r)\e_2dt_r}.
\end{equation}
In the NS limit, each column $\l_r$ contains an infinite number of boxes. We divide this infinite amount of boxes into infinitely many finite sets of size $k_{r,s}$, ranging from $u_{r,s}=\sum_{s'<s}k_{r,s'}$ to $u_{r,s}+k_{r,s}$:
\begin{equation}\label{before_mv}
\sum_{r=1}^{N_B}[f(u_r)-f(\bx_r)]=\sum_{r=1}^n\sum_{s=1}^\infty\int_{u_{r,s}}^{u_{r,s}+k_{r,s}}{f'(\bx_r+\e_2 t_{r,s})\e_2dt_{r,s}}=\e_2\sum_{r=1}^n\sum_{s=1}^\infty\int_{0}^{k_{r,s}}{f'(\tx_{r,s}+\e_2 t_{r,s})dt_{r,s}},
\end{equation}
with $\tx_{r,s}=\bar{x}_r+\e_2u_{r,s}$, $u_{r,0}=0$ and $u_{r,\infty}=\l_r$. Notations are displayed on figure \refOld{fig_d3}. By the mean value theorem, for each interval there exists $0\leq \bk_{r,s}\leq k_{r,s}$ such that
\begin{equation}\label{mean_value}
\int_0^{k_{r,s}}{f'(\tx_{r,s}+\e_2 t_{r,s})dt_{r,s}}=k_{r,s}f'(x_{r,s}),\quad x_{r,s}=\tx_{r,s}+\e_2\bk_{r,s}.
\end{equation}
The mean value $x_{r,s}$ depends on the function $f$. However, for functions of order $O(1)$, the variation of $x_{r,s}$ with $f$ is of order $O(\e_2)$ and thus can be neglected. It would not be the case if the $k_{r,s}$ were infinite, and indeed it is not possible to assign an average instanton position to a hadron with infinite charge because of its macroscopic size. Plugging \ref{mean_value} into \ref{before_mv}, we recover the relation \ref{id_ur_kx} for any test function $f(x)$ provided we identify $\{(k_{r,s},x_{r,s})\}\equiv\{(k_i,x_i)\}$. Thus, the relation between Bethe roots and hadronic variables corresponds to the decomposition of the Young tableaux columns into finite sets of $k_i$ boxes with an average instanton position $x_i$.

\subsubsection{Relation between bifundamental summands and integrands}
The formal relation we have observed between the variables $(k_i,x_i)$ and the Bethe roots enables the comparison between the integral expression \ref{CZ_A2} and the Bethe roots summation of the $A_2$ quiver partition function. The decomposition \ref{decomp_CZ} of the summands can be reproduced on the integrand $\mathscr{I}(k,x|l,y)$, with three parts that will be the analogue of the three bifundamental contributions $\CZ[M_1,Y_1]$, $\CZ[Y_1,Y_2]$ and $\CZ[Y_2,M_2]$ respectively:
\begin{align}
\begin{split}
\mathscr{I}(k,x|l,y)&=z_1(k,x)z_{12}(k,x|l,y)z_2(l,y)\quad\text{with:}\\
&z_1(k,x)=\prod_i\dfrac{m_1(x_i)^{k_i}}{(A_1(x_i)A_1(x_i+\e_1))^{k_i/2}}\exp\left(\frac{\e_2}{4}\sum_{i,j}k_ik_jG_{11}(x_i-x_j)\right)\\
&z_{12}(k,x|l,y)=\prod_i\dfrac{A_2(x_i+m+\e_1/2)^{k_i}}{(A_1(x_i)A_1(x_i+\e_1))^{k_i/2}}\prod_j\dfrac{A_1(y_j-m+\e_1/2)^{l_j}}{(A_2(y_j)A_2(y_j+\e_1))^{l_j/2}}\\
&\qquad\times \exp\left(\frac{\e_2}{4}\sum_{i,j}k_ik_jG_{11}(x_i-x_j)+\e_2\sum_{i,j}k_il_jG_{12}(x_i-y_j)+\frac{\e_2}{4}\sum_{i,j}l_il_jG_{22}(y_i-y_j)\right)\\
&z_2(l,y)=\prod_j\dfrac{m_2(y_j)^{l_j}}{(A_2(y_j)A_2(y_j+\e_1))^{l_j/2}}\exp\left(\frac{\e_2}{4}\sum_{i,j}l_il_jG_{22}(y_i-y_j)\right).
\end{split}
\end{align}
The factors $z_1(k,x)$ and $z_2(l,y)$ depend only on one set of variables and can be attached to the nodes $a=1,2$. On the other hand, $z_{12}(k,x|l,y)$ is related to the arrow $1\to2$ and will be referred as the \textit{bifundamental integrand}. Up to a pure mass term, the factor $z_1(k,x)$ can be obtained as $z_{12}(k,x|l',y')$ for a specific choice of variables $(l',y')$ such that
\begin{equation}
\e_2\sum_jl'_jG_{12}(z-y'_j)=\sum_f\log(z-m_f^{(1)}).
\end{equation}
These variables $(l',y')$ encode the information of the object $M_1$, in the same way that $(k,x)$ encodes $Y_1$. Similar ideas can also be found in \cite{Bourgine2012a} in the language of densities.

It is remarkable that performing the change of representation $(k,x)\to u_r$ (and $(l,y)\to v_s$), materialized by the relation \ref{id_densities_II} among densities, the bifundamental integrand $z_{12}(k,x|l,y)$ becomes $z[u,v]$, the NS limit of the bifundamental contribution \ref{zuv}. This formal identification may allow to define the KMZ transformation on the integrand expressions. In particular, the infinitesimal shifts of the Bethe roots $u_r\to u_r\pm\e_2$ correspond to a variation of the density $\rho_B(x)\to\rho_B(x)\mp\e_2\d'(x-u_r)$ which is rendered by a shift of $k_i\to k_i\pm1$ for the hadron of coordinate $x_i\simeq u_r+O(\e_2)$. The main difficulty lies in the decomposition of the function $\l(z)^2$ over poles, since in the representation $(k,x)$ this function is a product of essential singularities. It is still possible to use the Cauchy identity to write it as a sum over contributions at the location of the singularities, but the evaluation of the integrals is problematic.

On the other hand, the definition of $d_0(z)$, diagonal on the states $|k,x>=|u>$ easily follows from the definition \ref{def_D} of $D_0(z)$, and the limit \ref{dico_kx}, which gives, taking into account the $\e_2$ factor discrepancy in \ref{d0_series} between $d_0$ and $D_0$:
\begin{equation}\label{def_d0}
d_0(z)|k,x>=\e_2\sum_i\dfrac{k_i}{z-x_i}|k,x>.
\end{equation}
This expression can be integrated and expressed in terms of Bethe roots using \ref{id_densities_II}. It leads to the formula \ref{d0_ur} used in the previous section.

\section{Discussion}
In this paper, we considered the NS limit of the KMZ transformation that represents the action of the SHc generators on the bifundamental contribution to the $A_2$ quiver instanton partition function. In this limit, the bifundamental contribution can be written in terms of two sets of Bethe roots. We have defined a set of generators of degree $0,\pm1$ that exploits an infinitesimal variation of the Bethe roots. These generators reproduce the proper commutation relations, and generates an equivalent of the KMZ transformation. We then turned to an alternative approach that uses the Mayer cluster expansion to perform the NS limit. The coordinates and charges of the hadronic variables that appear in this formalism were formally related to the Bethe roots. This provides a hint on the possibility to realize the KMZ transformation directly on integral expressions. Finally, we studied a third expression of the partition function that takes the form of a Yang-Yang functional. Although obtained via Mayer expansion, this expression coincides with the limit of the Young tableaux summations under a proper identification of the Bethe roots. We deduced an action of the SHc generators on the functional (at first order in $\e_2$).

Now that the limit of the algebra has been identified, it remains to relate it to the algebraic structures of the underlying integrable systems, either the Yangian of XXX spin chains or the DDAHA of qNLS. The connection between SHc and the Calogero-Moser Hamiltonian for $\Nc=1$ allows to identify the states $|Y>$ with Jack polynomials. It is thus natural to expect a connection between the states $|u>$ and the eigenfunctions of the integrable systems. Such a connection could be obtained using the fact that the Calogero-Moser Cherednik operators reduce to the Dunkl operators of qNLS in a specific limit \cite{Bernard1993}. This limiting process may be the equivalent of the NS limit from the integrable point of view.

A better understanding of the states $|u>$ may also arise from the study of the single gauge group case ($A_1$ quiver). The connections between SHc and Gaiotto states \cite{Gaiotto2009b} has been discussed recently in \cite{Matsuo2014}. Some of these considerations may survive in the NS limit, and help to identify a semi-classical version of the Gaiotto states. We hope to address these issues in a near future.

\section*{Acknowledgements}
I would like to thank Davide Fioravanti for valuable comments at the early stage of this work. I am also indebted to Yutaka Matsuo for several fruitful discussions and many advices. It is a pleasure to acknowledge the Tokyo University (Hongo) for the kind hospitality and support during my stay there. I also want to thank Sergio Andraus Robayo for a very interesting discussion on Dunkl and intertwining operators. I acknowledge the Korea Ministry of Education, Science and Technology (MEST) for the support of the Young Scientist Training Program at the Asia Pacific Center for Theoretical Physics (APCTP).

\appendix
\section{Computing with instanton positions}\label{App0}
\subsection{Transition of notations}
To derive the dictionary between the notations of \cite{Kanno2013} and ours, let us drop for a moment the color and node indices, and focus on a single Young diagram. In \cite{Kanno2013}, Young diagrams were encoded as a sequence of $f$ rectangles characterized by the integers $0<r_1<\cdots<r_f$ and $0<s_f<\cdots<s_1$. It is further assumed that $r_0=s_{f+1}=0$. The transition between notations is done as follows: to each box $x\in A(Y)$ with $x=(l,i,j)$, corresponds an index $k\in\lbrackdbl1,f+1\rbrackdbl$ such that $i=r_{k-1}+1$ and $j=s_k+1$ (the labels $l$ and $a$ are implicit on $f,\ r_k,\ s_k,\cdots$). Defining $A_k(Y)=\b r_{k-1}-s_k-\xi_\text{KMZ}$ as in \cite{Kanno2013}, we deduce
\begin{equation}\label{transition_1}
a_{l,\text{KMZ}}+A_k(Y)=-(\phi_x+\e_+)/\e_2,
\end{equation}
where $a_{l,\text{KMZ}}$ denotes the Coulomb branch vevs in \cite{Kanno2013}, rescaled here as $a_{l,\text{KMZ}}=-a_l/\e_2$. In a similar way, for $x\in R(Y)$, there is a $k\in\lbrackdbl1,f\rbrackdbl$ such that $i=r_k$ and $j=s_k$, and defining $B_k(Y)=\b r_k-s_k$ we find
\begin{equation}\label{transition_2}
a_{l,\text{KMZ}}+B_k(Y)=-(\phi_x+\e_+)/\e_2,
\end{equation}
with the same rescaling of the Coulomb branch vevs.

\subsection{Useful formulas}
In some computations, it is not necessary to resort to the rectangle decomposition of Young tableaux. Here we provide some useful formulas in this respect. Considering the difference between sets of boxes that can be added or removed, we find
\begin{align}
\begin{split}
&\sum_{x\in A(Y)}\phi_x-\sum_{x\in R(Y)}\phi_x=\sum_{l=1}^{N_c}(a_l+\e_+f^{(l)}),\\
&\sum_{x\in A(Y)}\phi_x(\phi_x-\e_+)-\sum_{x\in R(Y)}\phi_x(\phi_x+\e_+)=\sum_la_l(a_l-\e_+)-2\e_1\e_2|Y|,
\end{split}
\end{align}
where $f^{(l)}$ is the number of rectangles in the Young diagram $Y^{(l)}$ and $|Y|$ the total number of boxes.

Let us also mention the identities coming from the expansion of \ref{prop_L} at $z\to\infty$:
\begin{align}
\begin{split}\label{prop_L2}
&\sum_{x\in A(Y)}\L_x(Y)^2-\sum_{x\in R(Y)}\L_x(Y)^2=\Nc,\\
&\sum_{x\in A(Y)}\phi_x\L_x(Y)^2-\sum_{x\in R(Y)}\phi_x\L_x(Y)^2=\sum_la_l+\hf\e_+\Nc(\Nc-1),\\
&\sum_{x\in A(Y)}\phi_x^2\L_x(Y)^2-\sum_{x\in R(Y)}\phi_x^2\L_x(Y)^2=-2\e_1\e_2|Y|+\sum_la_l^2+\e_+(\Nc-1)\sum_la_l+\dfrac16\e_+^2\Nc(\Nc-1)(\Nc-2).
\end{split}
\end{align}

The action of the commutators of two degree (minus) one operators on states $|Y>$ may be used to define some of the higher generators:
\begin{equation}\label{comm_D1D1}
[D_{\pm1}(z),D_{\pm1}(w)]|Y>=\sum_{x\in A/R(Y)}\sum_{y\in A/R(Y\pm x)}\dfrac{\L_x(Y)\L_y(Y\pm x)(z-w)(\phi_x-\phi_y)}{(z-\phi_x)(z-\phi_y)(w-\phi_x)(w-\phi_y)}|Y\pm x\pm y>.
\end{equation}

\section{Justifications}
\subsection{Order of $\L_x(Y)$}\label{AppB0}
Let $x\in A(Y)$ such that $x=(l,i,j)$ with $(i,j)\in Y^{(l)}$. The multiplicity of a column $\l_i^{(l)}$ of $Y^{(l)}$ is the number of columns in $Y^{(l)}$ with the same height $\l_i^{(l)}$. There is no other $y\in A(Y)$ in the same column, i.e. such that $\phi_y=\phi_x+O(\e_2)$. But there is always a unique $y\in R(Y)$ such that $\phi_y=\phi_x-\e_++O(\e_2)$ (except for $i=1$). There is a unique $y_1\in R(Y)$ with $\phi_{y_1}=\phi_x+O(\e_2)$, and a unique $y_2\in A(Y)$ with $\phi_{y_2}=\phi_x+\e_++O(\e_2)$, if and only if the column $\l_i^{(l)}$ is of multiplicity one. Thus, $\L_x(Y)=O(\sqrt{\e_2})$ except if $i=1$ for which $\L_x(Y)=O(1)$.

Now, we let $x=(l,i,j)\in R(Y)$. There is no other $y\in R(Y)$ with $\phi_y=\phi_x+O(\e_2)$, but there is always a unique $y\in A(Y)$ such that $\phi_y=\phi_x+\e_++O(\e_2)$. There is a unique $y\in A(Y)$ such that $\phi_y=\phi_x+O(\e_2)$ if and only if $\l_i^{(l)}$ is of multiplicity one. Likewise, there is a unique $y\in R(Y)$ such that $\phi_y=\phi_x-\e_++O(\e_2)$ if and only if $\l_i^{(l)}$ is of multiplicity one, and $i>1$. Thus, if $i\neq 1$, we have $\L_x(Y)=O(\sqrt{\e_2})$. If $i=1$, $\l_1^{(l)}$ is of multiplicity one, and $\L_x(Y)=O(1)$.

\subsection{NS limit of the bifundamental contribution}\label{App_BZ}
\begin{figure}[!t]
\centering
\includegraphics[width=4cm]{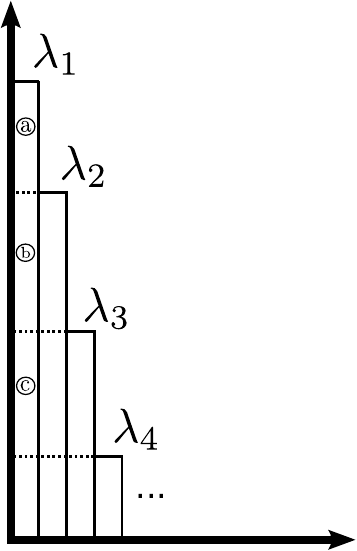}
\caption{Young tableaux with a labeling of boxes in the first column (we dropped the node and color indices).}
\label{fig_d4}
\end{figure}

In this appendix, we take the NS limit of the bifundamental contribution \ref{CZ_YY}. A similar calculation can be found in \cite{Chen2011}. The cut-offs $n_l^{(a)}$ are assumed to be all equal to $L$ which will be sent to infinity at the end of the computation. We first concentrate on a single object $Y$, and examine the following quantity that is involved in the denominator:
\begin{equation}
P(x)=\prod_{(l,i,j)\in Y}\prod_{l'=1}^{\Nc}(\tilde{t}_{l,j}-t_{l',i}+x).
\end{equation}
In the NS limit, $t_{l',i}$ is identified with the Bethe root $u_r$, and the main difficulty is to treat the dual quantities $\tilde{t}_{l,j}$. For this purpose, let us focus on the $l$th Young diagram pictured in figure \refOld{fig_d4}, and consider the first column. Boxes in the set denoted $(a)$ are such that $\tilde{t}_{l,j}=a_l+\e_1+(j-1)\e_2$ for $j=\l_2^{(l)}+1\cdots \l_1^{(l)}$. Similarly, boxes in the set $(b)$ are mapped to $\tt_{l,j}=a_l+2\e_1+(j-1)\e_2$ with $j=\l_3^{(l)}+1\cdots \l_2^{(l)}$. In general, the first column should be divided into $n_l=L$ sets of boxes with $\tt_{l,j}=a_l+r\e_1+(j-1)\e_2$, $j=\l_{r+1}^{(l)}+1\cdots \l_r^{(l)}$ with $r=1\cdots L$ and $\l_{L+1}^{(l)}=0$. The same applies to the next columns, but we should start at $r=i$ for the $i$th column. We thus have found
\begin{equation}
P(x)=\prod_{l,l'=1}^{\Nc}\prod_{i=1}^L\prod_{r=i}^L\prod_{j=\l_{r+1}^{(l)}+1}^{\l_r^{(l)}}(\tt_{l,j}-t_{l',i}+x).
\end{equation}
Plugging in the explicit expression for $\tt_{l,j}$, we obtain
\begin{equation}
P(x)=\prod_{l,l'=1}^{\Nc}\prod_{i=1}^L\prod_{r=i}^L\e_2^{\l_r^{(l)}-\l_{r+1}^{(l)}}\!\!\!\!\prod_{j=\l_{r+1}^{(l)}+1}^{\l_r^{(l)}}(y_{ll'ir}+(j-1)),\quad \e_2y_{ll'ir}=a_l+r\e_1+x-t_{l,i}.
\end{equation}
The product over $j$ can now be replaced by a ratio of gamma functions,
\begin{equation}
P(x)=\prod_{l,l'=1}^{\Nc}\prod_{i=1}^L\prod_{r=i}^L\e_2^{\l_r^{(l)}-\l_{r+1}^{(l)}}\dfrac{\G[y_{ll'ir}+\l_r^{(l)}]}{\G[y_{ll'ir}+\l_{r+1}^{(l)}]}.
\end{equation}
As $\e_2\to0$, the gamma functions arguments tend to infinity, and we can use the Stirling approximation $\G[x]\simeq (x/e)^x$ (the square root is subleading). Noticing that $\e_2(y_{ll'ir}+\l_r^{(l)})=t_{l,r}-t_{l',i}+x+\e_1$ and $\e_2(y_{ll'ir}+\l_r^{(l)})=t_{l,r+1}-t_{l',i}+x$, and introducing the function $g(x)=x^{x/\e_2}$, we can write
\begin{equation}
P(x)\simeq\prod_{l,l'}^{\Nc}\prod_{\superp{r,i=1}{r\geq i}}^{L}e^{\l_{r+1}^{(l)}-\l_r^{(l)}}\dfrac{g(t_{l,r}-t_{l',i}+x+\e_1)}{g(t_{l,r+1}-t_{l',i}+x)}.
\end{equation}
This expression now involves only the variables $t_{l,i}$ which can be replaced by Bethe roots. It is however better to slightly shift the products indices to get
\begin{equation}
P(x)\simeq e^{-\Nc |Y|}\prod_{l,l'}^{\Nc}\prod_{i=1}^L\dfrac{g(t_{l,i}-t_{l',i}+x+\e_1)}{g(t_{l,L+1}-t_{l',i}+x)}\times\prod_{l,l'=1}^{\Nc}\prod_{\superp{r,i=1}{r>i}}^{L}\dfrac{g(t_{l,r}-t_{l',i}+x+\e_1)}{g(t_{l,r}-t_{l',i}+x)}.
\end{equation}

The previous result upon $P(x)$ can be used to take the NS limit of the denominator factors in \ref{CZ_YY}. Considering
\begin{equation}
Q^2=\prod_{(l,i,j)\in Y}\prod_{l'=1}^{\Nc}(\tilde{t}_{l,j}-t_{l',i}+\e_2)(\tilde{t}_{l,j}-t_{l',i}-\e_1),
\end{equation}
we find
\begin{equation}
Q^2\simeq e^{-2\Nc Y}\prod_{l,l'}^{\Nc}\prod_{i=1}^L\dfrac{g(t_{l,i}-t_{l',i})g(t_{l,i}-t_{l',i}+\e_1)}{g(t_{l,L+1}-t_{l',i})g(t_{l,L+1}-t_{l',i}-\e_1)}
\times\prod_{l,l'=1}^{\Nc}\prod_{\superp{r,i=1}{r>i}}^{L}\dfrac{g(t_{l,r}-t_{l',i}+\e_1)}{g(t_{l,r}-t_{l',i}-\e_1)}.
\end{equation}
It is possible to use the property $g(e^{\pm i\pi}x)g(x)=e^{\pm i\pi x/\e_2}$ to symmetrize the product. An ambiguity arises in the choice of the sense of the rotation. It is a part of the overall ill-definiteness for the final result of $z[u,v]$. Here, we make the choices that provide the simplest expressions: we rotate both numerator and denominators in the ratios in the same direction, $e^{+i\pi}$. After re-arranging the diagonal terms $r=i$, we obtain
\begin{equation}
Q^2\simeq e^{-2\Nc Y}e^{i\pi\Nc^2L^2\e_1/2\e_2}\prod_{l,l'}^{\Nc}\prod_{i=1}^Lg(t_{l',i}-t_{l,L+1})g(t_{l',i}-t_{l,L+1}+\e_1)\times\prod_{l,l'=1}^{\Nc}\prod_{r,i=1}^{L}\left(\dfrac{g(t_{l,r}-t_{l',i}+\e_1)}{g(t_{l,r}-t_{l',i}-\e_1)}\right)^{1/2}.
\end{equation}
Finally, replacing $t_{l,i}\to u_r$ and $t_{l,L+1}\to\xi_l$ ($N_B=\Nc L$), we find:
\begin{equation}
Q^2\simeq e^{-2\Nc |Y|}e^{i\pi\NB^2\e_1/2\e_2}\prod_{l=1}^{\Nc}\prod_{r=1}^{N_B}g(u_r-\xi_l)g(u_r-\xi_l+\e_1)\times\prod_{r,s=1}^{N_B}\left(\dfrac{g(u_r-u_s+\e_1)}{g(u_r-u_s-\e_1)}\right)^{1/2}.
\end{equation}

A similar treatment can be applied to the numerator, but we now have to keep track of the node indices. This does not add much difficulty, and we find for the numerator of \ref{CZ_YY} the expression
\begin{align}
\begin{split}
&e^{-\Nc |Y_1|-\Nc |Y_2|}e^{i\pi\NB^2\e_1/2\e_2}\prod_{l=1}^{\Nc}\prod_{r=1}^{N_B^{(1)}}g(u_r-\xi_l^{(2)}+m+\e_1/2)\\
\times&\prod_{l=1}^{\Nc}\prod_{r=1}^{N_B^{(2)}}g(v_r-\xi_l^{(1)}-m+\e_1/2)\times\prod_{r=1}^{N_B^{(1)}}\prod_{s=1}^{N_B^{(2)}}\dfrac{g(u_r-v_s+m+\e_1/2)}{g(u_r-v_s+m-\e_1/2)}.
\end{split}
\end{align}
Taking the ratio, we end up with the formula given in \ref{zuv}.

\section{Nekrasov-Shatashvili limit from Mayer expansion}\label{AppA}
\subsection{Equations of motion, grand-canonical free energy}\label{AppA1}
In this appendix, we study a general class of grand-canonical partition functions defined as the discrete Laplace transform
\begin{equation}\label{defZGC}
\ZGC(q_1,\cdots,q_M)=\sum_{N_1,\cdots,N_M=0}^\infty{\prod_{a=1}^M\dfrac{(q_a/\e)^{N_a}}{N_a!}\ \ZC(N_1,\cdots N_M)}
\end{equation}
of the following canonical partition functions
\begin{equation}\label{defZC}
\ZC(N_1,\cdots N_M)=\int{\prod_{\superp{a,b=1}{a<b}}^M\prod_{i=1}^{N_a}\prod_{j=1}^{N_b}K_{ab}(\phi_{i,a}-\phi_{j,b})\prod_{a=1}^M\left(\prod_{\superp{i,j=1}{i<j}}^{N_a}K_{aa}(\phi_{i,a}-\phi_{j,a})\prod_{i=1}^{N_a}Q_a(\phi_{i,a})\dfrac{d\phi_{i,a}}{2i\pi}\right)}.
\end{equation}
Partition functions with $M=1$ were considered in \cite{Bourgine2014}, and we show here that the method extends smoothly to $M>1$. The integration contour is along the real axis, closed in the upper half plane, and avoiding possible singularities at infinity. We study the limit $\e\to0$, with the kernels defined as
\begin{equation}
K_{ab}(x)=1+\e G_{ab}(x)+\e\d_{ab}p_a(x),\quad p_a(x)=\dfrac{\a_a\e}{x^2-\e^2},
\end{equation}
and $G_{ab}(x)$ a function independent of $\e$. Note that the poles at $x=\pm\e$ that pinch the integration contour are only present in the diagonal term $K_{aa}$, the non-diagonal kernel $K_{ab}$ with $a\neq b$ is not singular at $\e\to0$. This is in agreement with quivers instanton partition functions where $K_{ab}$ corresponds to the bifundamental contribution (with a non-zero mass $m_{ab}$), and $K_{aa}$ to the vector hypermultiplet.\footnote{For the application to the $A_2$ quiver partition function, we take $K_{aa}(x)=\D(x)\D(-x)$ and $K_{ab}(x)=1/\D(x+m_{ab}-\e_+/2)$ with $\D(x)=x(x+\e_+)/((x+\e_1)(x+\e_2))$ \cite{Shadchin2005}. It gives in the NS limit with $\e=\e_2$, $K_{aa}(x)=1+\e G(x)+\e p_a(x)+O(\e^2)$ (with $\a_a=1$) and $K_{12}(x)=1+\e G_{12}(x)+O(\e^2)$. The functions $G(x)$ and $G_{12}(x)$ are defined in \ref{G11}.} We will further assume that $G_{ab}(x)$ satisfies the property  $G_{ab}(x)=G_{ba}(-x)$. This condition is necessary to be able to expand over non-oriented clusters.\footnote{It corresponds to the condition $S_{ab}(x)S_{ba}(-x)=1$ for the scattering amplitudes, $G_{ab}(x)=-\p_x\log S_{ab}(x)$, and can be obtained as a consequence of unitarity and hermitian analyticity.} In particular, this implies that the diagonal entries $G_{aa}(x)$ are even functions of $x$.

It is possible to express the integrand of the partition function \ref{defZC} with the help of double indices $I=(i,a)$ (and $J=(j,b)$) for $a=1\cdots M$ and $i=1\cdots N_a$, and with the lexicographical order $I<J$ if and only if $a<b$ or ($a=b$ and $i<j$). Applying the Mayer expansion technique \cite{Mayer1940,Mayer1941,Andersen1977} to \ref{defZGC} for $K_{ab}=1+\e f_{ab}$, we find for the free energy $\FGC=\e\log\ZGC$:
\begin{equation}
\FGC=\sum_{l_1,\cdots l_M=0}^\infty{\sum_{\bC_l}\dfrac{\e^{-(l-1)}}{\s(\bC_l)}\int{\prod_{I\in V(\bC_l)}q_aQ_a(\phi_I)\dfrac{d\phi_I}{2i\pi}\prod_{<IJ>\in E(\bC_l)}\e f_{ab}(\phi_I-\phi_J)}}.
\end{equation}
The connected clusters $\bC_l$ are made of $l=\sum_al_a$ vertices $I=(i,a)$. We denote by $V(\bC_l)$ the set of vertices of the cluster $\bC_l$, and $E(\bC_l)$ the set of links $<IJ>$ connecting two vertices $I$ and $J$. Each vertex bear a color label $a$, and there are $l_a$ vertices of type $a$.\footnote{Let us emphasize that the color of the vertices corresponds to the node index $a$ of the quiver. It has nothing to do with the gauge color index of fields in the gauge theory.} The symmetry factor $\s(\bC_l)$ of the cluster is the cardinal of the group of automorphisms preserving the cluster. These automorphisms must also preserve the color labeling because the combinatorial factor in the definition of $\ZGC$ is $\prod_a N_a!$ and not $(\sum_a N_a)!$ as in the usual Mayer cluster expansion. Decomposing $f_{ab}=G_{ab}+\d_{ab}p_a$, the cluster expansion becomes a sum over clusters $C_l$ with two types of links: G-links with kernel $G_{ab}$ between vertices of color $a$ and $b$, and p-links of kernel $p_a$ between two vertices of the same color $a$:
\begin{equation}
\FGC=\sum_{l_1,\cdots l_M=0}^\infty{\sum_{C_l}\dfrac{\e^{-(l-1)}}{\s(C_l)}\int{\prod_{I\in V(\bC_l)}q_aQ_a(\phi_I)\dfrac{d\phi_I}{2i\pi}\prod_{<IJ>\in E_p(C_l)}\dfrac{\a_a\e^2}{(\phi_I-\phi_J)^2-\e^2}\prod_{<IJ>\in E_G(C_l)}\e G_{ab}(\phi_I-\phi_J)}}.
\end{equation}
We have $E(C_l)=E_p(C_l)\cup E_G(C_l)$ where $E_p$ denotes the set of p-links, and $E_G$ the set of G-links. The clusters that contribute at first order in $\e$ have no cycle involving $G$-links, but do have cycles of only p-links. They are of order $O(\e^{l-1})$ so that $\FGC$ defined above is of order one \cite{Bourgine2014}.

The generating function of rooted vertices, also called dressed vertex, bears an index $a$ corresponding to the color of the root,
\begin{equation}\label{cluster_Y}
Y^{(a)}(x)=q_aQ_a(x)\sum_{l_1,\cdots l_M=0}^\infty{\sum_{C_l^{x,a}}\dfrac{\e^{-(l-1)}}{\s(C_l^{x,a})}\int{\prod_{\superp{I\in V(\bC_l^{x,a})}{I\neq x}}q_aQ_a(\phi_I)\dfrac{d\phi_I}{2i\pi}\!\!\!\!\prod_{<IJ>\in E_p(C_l^{x,a})}\!\!\!\!\e p_a(\phi_{IJ})\!\!\!\!\prod_{<IJ>\in E_G(C_l^{x,a})}\!\!\!\!\e G_{ab}(\phi_{IJ})}}.
\end{equation}
where we denoted $\phi_{IJ}=\phi_I-\phi_J$. The summation is over rooted clusters $C_l^{x,a}$, with the root $x=\phi_x$ of color $a$. The symmetry factor $\s(C_l^{x,a})\leq\s(C_l)$ is the number of automorphisms of $C_l$ that leave the root invariant. The generating functions $Y_G^{(a)}(x)$ and $q_aQ_a(x)Y_p^{(a)}(x)$ are defined with the additional requirement that the root $x$ is linked to other vertices with only G- or p- links respectively. At first order, we have the factorization property
\begin{equation}
Y^{(a)}(x)\simeq Y_G^{(a)}(x)Y_p^{(a)}(x),
\end{equation}
as a consequence of the property that G-links do not form cycles. 

Minimal clusters that contribute to $Y_G^{(a)}$ can be decomposed into sub-clusters rooted by the direct descendants. The root $x$, of type $a$, is linked to $m_b$ vertices $\phi_{j,b}$ of type $b$ through a link $G_{ab}$, which leads to
\begin{equation}
Y_G^{(a)}(x)=q_aQ_a(x)\sum_{m_1,\cdots,m_M=0}^\infty\prod_{b=1}^M\dfrac{\e^{-m_b}}{m_b!}\prod_{i=1}^{m_b}\int{\e G_{ab}(x-\phi_{i,b})Y^{(b)}(\phi_{i,b})\dfrac{d\phi_{i,b}}{2i\pi}}.
\end{equation}
The symmetry factors $m_b!$ take into account the possibility of permuting vertices of the same color. Contributions of sub-clusters factorizes to give
\begin{equation}\label{eom1}
Y_G^{(a)}(x)=q_aQ_a(x)\exp\left(\sum_{b=1}^M\int{G_{ab}(x-y)Y^{(b)}(y)\dfrac{dy}{2i\pi}}\right).
\end{equation}
This is the generalization of the formula (2.6) of \cite{Bourgine2014}.

The third relation is obtained by repeating the confinement argument employed in \cite{Bourgine2014}. This argument remains unchanged because p-links only relate vertices of the same color. It gives the following relation between $Y^{(a)}(x)$ and $Y_G^{(a)}(x)$:
\begin{equation}\label{eom2}
Y^{(a)}(x)=l_{\a_a}(Y_G^{(a)}(x)),
\end{equation}
with the function $l_\a(x)$ defined in equ (2.13) of \cite{Bourgine2014}. For $\a=1$, this function simplifies into a logarithm: $l_1(x)=-\log(1-x)$.

The fields $\rho_a$ and $\vphi_a$ are introduced as
\begin{equation}\label{def_rho_phi}
Y^{(a)}(x)=2i\pi\rho_a(x),\quad Y_G^{(a)}(x)=q_aQ_a(x)e^{-\vphi_a(x)}\implies 2i\pi\rho_a(x)=l_{\a_a}(q_aQ_a(x)e^{-\vphi_a(x)}).
\end{equation}
Exploiting the relations \ref{eom1} and \ref{eom2}, we obtain the TBA-like NLIE:
\begin{equation}\label{eom_rhophi}
\vphi_a(x)+\sum_b\int{G_{ab}(x-y)l_{\a_b}(q_bQ_b(y)e^{-\vphi_b(y)})\dfrac{dy}{2i\pi}}=0.
\end{equation}
For $\a_a=1$, we recover the NLIE \ref{NLIE_quiver} with the pseudo-energies $\ve_a$ defined as $\ve_a(x)=\vphi_a(x)-\log Q_a(x)$.

\paragraph{Back to the free energy:} The grand-canonical (instanton) free energy is obtained from the technique displayed in \cite{Bourgine2014} simply by introducing an additional sum over the different colors. The expression of $\FGC$ follows from the Basso-Sever-Vieira formula, $\FGC=\G_0-(1/2)\G_1$ with
\begin{equation}
\G_0=\sum_{a=1}^M\int{\dfrac{dx}{2i\pi}L_{\a_a}(Y_G^{(a)}(x))},\quad \G_1=\sum_{a,b=1}^M\int{Y^{(a)}(x)Y^{(b)}(y)G_{ab}(x-y)\dfrac{dxdy}{(2i\pi)^2}}.
\end{equation}
The function $L_\a(x)$ is a primitive of $l_\a(x)$, it reduces at $\a=1$ to the dilogarithm $L_1(x)=\text{Li}_2(x)$. Plugging in the equations of motion, and using the variable $\rho_a$ and $\vphi_a$ instead of $Y^{(a)}$ and $Y_G^{(a)}$, we recover the results of \cite{Meneghelli2013}: $\FGC=\SGC[\rho_a,\vphi_a]$ on-shell with the effective action
\begin{equation}\label{SGC_onshell}
\SGC[\rho_a,\vphi_a]=\hf\sum_{a,b=1}^M\int{\rho_a(x)\rho_b(y)G_{ab}(x-y)dxdy}+\sum_{a=1}^M\int{\rho_a(x)\vphi_a(x)dx}+\dfrac1{2i\pi}\sum_{a=1}^M\int{L_{\a_a}(q_aQ_a(x)e^{-\vphi_a(x)})dx}.
\end{equation}
The equations of motion reproduce the relations \ref{def_rho_phi} and \ref{eom_rhophi}.

\subsection{Confinement}\label{AppA2}
For simplicity, in this section we focus on the case $M=2$, but the generalization to higher $M$ is straightforward. To obtain the hadronic partition function, we mimic the treatment performed in \cite{Bourgine2014}. At first order in $\e_2$, the instanton partition function can be written as a path integral,
\begin{equation}
\ZGC(q_1,q_2)\simeq\int{D[\rho_a,\vphi_a]e^{\frac1\e\SGC[\rho_a,\vphi_a]}}.
\end{equation}
The function $L_{\a_a}$ contains all the dependence in $q_a$, expanding it we get
\begin{equation}
\exp\left(\dfrac1\e\int{L_{\a_a}(q_aQ_a(x)e^{-\vphi_a(x)})\dfrac{dx}{2i\pi}}\right)=\sum_{p=1}^\infty\dfrac{\e^{-p}}{p!}\sum_{k_1,k_2,\cdots, k_p=1}^\infty\int{\prod_{i=1}^pI_{k_i}^{(a)}q_a^{k_i}Q_a(x_i)^{k_i}e^{-k_i\vphi_a(x_i)}\dfrac{dx_i}{2i\pi}},
\end{equation}
with $I_k^{(a)}$ the coefficients of the Taylor expansion of $L_{\a_a}(x)$:
\begin{equation}
L_{\a_a}(x)=\sum_{k=1}^\infty I_k^{(a)}x^k.
\end{equation}
Plugging this expansion in the expression of $\ZGC(q_1,q_2)$, we find
\begin{align}
\begin{split}
\ZGC(q_1,q_2)\simeq&\sum_{p,q=1}^\infty\dfrac{\e^{-p-q}}{p!q!}\sum_{\{k_i\}_{i=1}^p}\sum_{\{l_j\}_{j=1}^q}\int\prod_{i=1}^pI_{k_i}^{(1)}q_1^{k_i}Q_1(x_i)^{k_i}\dfrac{dx_i}{2i\pi}\prod_{j=1}^qI_{l_j}^{(2)}q_2^{l_j}Q_2(y_j)^{l_j}\dfrac{dy_j}{2i\pi}\\
&\int D[\rho_a,\vphi_a]e^{\frac1{2\e}\sum_{a,b}\int{\rho_a(x)\rho_b(y)G_{ab}(x-y)dxdy}}e^{\frac1{\e}\int{\vphi_1(x)\left(\rho_1(x)-\e\sum_ik_i\d(x-x_i)\right)dx}}\\
&\times e^{\frac1{\e}\int{\vphi_2(y)\left(\rho_2(y)-\e\sum_jl_j\d(y-y_j)\right)dy}}.
\end{split}
\end{align}
The fields $\vphi_a$ appear as Lagrange multipliers enforcing the conditions \ref{cond_density} for the densities (with $\e=\e_2$). Replacing the densities, we find
\begin{align}
\begin{split}\label{cluster_hadron}
\ZGC(q_1,q_2)\simeq&\sum_{p,q=1}^\infty\dfrac{\e^{-p-q}}{p!q!}\sum_{\{k_i\}_{i=1}^p}\sum_{\{l_j\}_{j=1}^q}\int\prod_{i=1}^pI_{k_i}^{(1)}q_1^{k_i}Q_1(x_i)^{k_i}\dfrac{dx_i}{2i\pi}\prod_{j=1}^qI_{l_j}^{(2)}q_2^{l_j}Q_2(y_j)^{l_j}\dfrac{dy_j}{2i\pi}\\
&\times \exp\left(\frac{\e}2\sum_{i,j}k_ik_jG_{11}(x_i-x_j)+\frac{\e}2\sum_{i,j}l_il_jG_{11}(y_i-y_j)+\e\sum_{i,j}k_il_jG_{12}(x_i-y_j)\right).
\end{split}
\end{align}
The expression given in \ref{CZ_A2} is obtained after specialization of \ref{defZGC} to the quiver partition function \ref{def_CZA2}. It implies to set $\e=\e_2$, $\a_1=\a_2=1$ leading to $I_k^{(a)}=1/k^2$. The kernels $G_{11}$ and $G_{22}$ coincide with the function $G$ given in \ref{G11}.

\bibliographystyle{hunsrt.bst}

\end{document}